\documentclass[twocolumn,twocolappendix,trackchanges]{aastex701}	% linenumbers,
\usepackage[mathlines]{lineno}	% For arXiv

\shorttitle{Fast {\sc Imcom} in 1D}
\shortauthors{K. Cao}
\usepackage{CJKutf8}

\usepackage{graphicx}	% Including figure files
\usepackage{amsmath}	% Advanced maths commands
\usepackage{amssymb}	% Extra maths symbols
\usepackage{multirow}

\usepackage[OT2,T1]{fontenc}
\DeclareSymbolFont{cyrletters}{OT2}{wncyr}{m}{n}
\DeclareMathSymbol{\Sha}{\mathalpha}{cyrletters}{"58}

\newcommand{\papone}{Paper I}
\newcommand{\paptwo}{Paper II}
\newcommand{\papthree}{Paper III}
\newcommand{\papfour}{Paper IV}

\begin{document}

\title{Linear Image Regridding and Coaddition with Oversampled Point Spread Functions: Lessons from 1D}

\author[orcid=0000-0002-1699-6944]{Kaili Cao (\begin{CJK*}{UTF8}{gbsn}曹开力\end{CJK*})}
\affiliation{Center for Cosmology and AstroParticle Physics (CCAPP), The Ohio State University, 191 West Woodruff Ave, Columbus, OH 43210, USA}
\affiliation{Department of Physics, The Ohio State University, 191 West Woodruff Ave, Columbus, OH 43210, USA}
\email[show]{cao.1191@osu.edu}

\collaboration{all}{Roman HLIS Cosmology PIT}

\begin{abstract}

Image regridding and coaddition have a wide range of applications in astronomical observations. {\sc Imcom}, an algorithm that provides control over point spread function (PSF) and noise in coadded images, has been found to meet the stringent requirements of weak gravitational lensing cosmology with the forthcoming Nancy Grace Roman Space Telescope. In this work, I introduce a new algorithm, Fast {\sc Imcom}, which outperforms traditional {\sc Imcom} in terms of both efficiency and quality. After explaining the underlying philosophy and mathematical formalism, I conduct systematic comparisons between {\sc Imcom} and Fast {\sc Imcom} in terms of PSF reconstruction in 1D. While a 2D implementation is beyond the scope of this paper, I demonstrate how to generalize Fast {\sc Imcom} to 2D and discuss practical issues involved. This new algorithm has the potential of reducing both the computational costs and storage requirements (current estimates are $\sim 100\,{\rm M}$ core-hours and $\sim 1.5 \,{\rm PB}$, respectively) of the Roman High Latitude Imaging Survey (HLIS) by an order of magnitude. Meanwhile, it provides implications for the dithering patterns of Roman surveys (extrapolated from 1D to 2D). I also address potential applications of Fast {\sc Imcom} beyond the Roman HLIS, with focus on other weak lensing programs and Roman time domain surveys; the actual range of use cases is likely beyond what is discussed here.

\end{abstract}

	% https://astrothesaurus.org

\section{Introduction} \label{sec:intro}

Image regridding and coaddition are common tasks in astronomical image processing. For deep surveys like Hubble Deep Fields \citep{1996AJ....112.1335W, 2000ARA&A..38..667F, 2006AJ....132.1729B}, images are stacked to achieve greater depths. For time domain missions like Kepler and K2 \citep{2010Sci...327..977B, 2010ApJ...713L..79K, 2014PASP..126..398H}, references images (also known as ``templates'' in some fields of study) are indispensable for conducting difference image analysis. For weak gravitational lensing cosmology \citep[see][for some recent reviews]{2013PhR...530...87W, 2015RPPh...78h6901K, 2018ARA&A..56..393M}, oversampled images are necessary for accurate measurements of galaxy shapes.

Such image processing procedures are usually formulated as linear transformations so that the output images have well-defined point spread functions \citep[PSFs;][]{2023OJAp....6E...5M}. \citet{1999PASP..111..227L} demonstrated that ``superimages'' with Nyquist sampling can be reliably constructed from undersampled dithered images in Fourier space under some circumstances. To handle arbitrary (both translational and rotational) dithers of images, {\sc Drizzle} \citep{2002PASP..114..144F, 2012drzp.book.....G} has been widely used for a few decades. While {\sc Drizzle} is robust and efficient, its output images lack well-defined PSFs and noise fields. A relatively new algorithm, {\sc Imcom} \citep[``IMage COMbination'';][]{2011ApJ...741...46R}, overcomes these issues by leveraging our knowledge about PSFs in native images.

In a series of papers, my co-authors and I have applied the {\sc Imcom} algorithm to simulated images of the forthcoming Nancy Grace Roman Space Telescope \citep[hereafter Roman;][]{2019arXiv190205569A, 2025arXiv250510574O}. In \citet[][hereafter \papone]{2024MNRAS.528.2533H}, we stated the problem of image coaddition in a modern context and presented first results of coadding images simulated by \citet{2023MNRAS.522.2801T}. In \citet[][hereafter \paptwo]{2024MNRAS.528.6680Y}, we analyzed noise properties and point sources in these results and found that {\sc Imcom} meets requirements of the Roman weak lensing program. In \citet[][hereafter \papthree]{2025ApJS..277...55C}, we reorganized the program and introduced new linear algebra strategies to enhance comprehensibility and efficiency. In \citet[][hereafter \papfour]{2025arXiv250918286C}, we systematically explored the impact of {\sc Imcom} hyperparameters and found that some configurations, especially wider Gaussian target output PSFs, lead to better coadded images.

Although {\sc Imcom} has been successful and gradually become more mature, its widespread usage is still hindered by its non-ideal computational expense. Meanwhile, {\sc Imcom} leaves some minor but undesirable artifacts (e.g., postage stamp boundary effects; see \papthree\ for discussion) in the output images, and some of its behaviors (e.g., why the target PSF width affects diagnostics in the manner observed in \papfour) have not been thoroughly understood. In this work, I introduce a new algorithm, Fast {\sc Imcom}, to address these difficulties and showcase its performance on PSF reconstruction in 1D. This paper is structured as follows. Section~\ref{sec:essay} is an essay on the underlying philosophy of PSF manipulation. On this basis, I present the mathematical formalisms of {\sc Imcom} and Fast {\sc Imcom} in Section~\ref{sec:meth}. Then in Sections~\ref{sec:regrid} and \ref{sec:coadd}, I investigate image regridding and coaddition in 1D, respectively, and describe how to generalize Fast {\sc Imcom} to 2D. Multiple technical aspects and potential scientific applications of the envisioned 2D implementation are discussed in Section~\ref{sec:disc}, before major findings and implications are summarized in Section~\ref{sec:summ}. In Appendix~\ref{app:asym}, I study the impact of asymmetric windows for input pixels, which are very common with the Cholesky kernel of {\sc Imcom} (introduced in \papthree).

\section{An Essay on PSF} \label{sec:essay}

To lay the foundation for the following sections, I start by pondering fundamental ingredients of image regridding and coaddition --- definitions of PSFs, functions being sampled, and information content of images --- in Sections~\ref{ss:psf_def}, \ref{ss:samp}, and \ref{ss:info}, respectively. Then I introduce 1D PSFs used in this work in Section~\ref{ss:psf_1d}.

\subsection{What is a point spread function?} \label{ss:psf_def}

As its name indicates, a PSF describes how the light from a point source is spread over an imaging device. To quote a classical text, \citet{1987PASP...99..191S} defined PSF as ``the two-dimensional brightness distribution produced in the detector by the image of an unresolved source, such as a star.'' Mathematically, a monochromatic PSF is usually formulated as a normalized function $({\mathbb R}^2, {\mathbb R}^2) \mapsto {\mathbb R}^+$:\footnote{${\mathbb R}$ denotes the set of real numbers, ${\mathbb R}^2$ denotes the set of $2$-tuples of real numbers (e.g., coordinates of points on a 2D Euclidean plane), and ${\mathbb R}^+$ denotes the set of non-negative real numbers.} $G ({\boldsymbol s}; {\boldsymbol r})$, where ${\boldsymbol r}$ is the position of the point source, and ${\boldsymbol s}$ is the relative position of the point on the imaging device.\footnote{In {\sc Imcom} papers, the relative position is more explicitly written as ${\boldsymbol r} - {\boldsymbol s}$ or ${\boldsymbol r} - {\boldsymbol r}'$. The physical meaning of $G$ is the same.} If the spatial variation of the PSF can be ignored, the function can be simplified to ${\mathbb R}^2 \mapsto {\mathbb R}^+$: $G ({\boldsymbol s})$. For a given optical system, the function $G$ also depends on the spectral energy distribution of the source. For simplicity, I only consider monochromatic PSFs in this work; handling of chromaticity is discussed in Section~\ref{ss:tech}.

Imaging devices usually consist of regular arrays of pixels. Therefore, if we think of a PSF as the probability distribution of the landing location of a photon from a known direction, it should be formulated as $({\mathbb Z}^2, {\mathbb R}^2) \mapsto {\mathbb R}^+$:\footnote{${\mathbb Z}$ denotes the set of real integers, and ${\mathbb Z}^2$ denotes the set of $2$-tuples of real integers (e.g., coordinates of points on a 2D integer lattice).} $G' ({\boldsymbol i}; {\boldsymbol r})$, where ${\boldsymbol i}$ is the pixel index. Note that ${\boldsymbol i}$ is written as a vector so that the expression of $G'$ is universal for all spatial dimensions; in 2D (and above), ${\boldsymbol i}$ may need to be flattened to facilitate operations in computers. Unlike the unpixelated $G$, the pixelated $G'$ necessarily depends on the position of the source ${\boldsymbol r}$: Even if $G$ only depends on the relative position ${\boldsymbol s}$, $G'$ still depends on where ${\boldsymbol r}$ is within the central pixel.

To relate $G$ and $G'$, here I define the pixelation function (following the convention of considering the center of a pixel as its position) as
\begin{equation}
    \Pi ({\boldsymbol s}; {\bf D}) = \Theta (1/2 - \Vert {\bf D} {\boldsymbol s} \Vert_\infty),
    \label{eq:pixelation}
\end{equation}
where the matrix ${\bf D}$ characterizes the (linear) distortion of the pixel array, $\Theta (\cdot)$ is the Heaviside step function, and $\Vert \cdot \Vert_\infty$ is the $L$-infinity norm. The sampling Dirac comb is defined as
\begin{equation}
    \Sha ({\boldsymbol s}; {\boldsymbol s}_0, {\bf D}) = \sum_{{\boldsymbol i} \in {\mathbb Z}^2} \delta^2 ({\bf D} ({\boldsymbol s} - {\boldsymbol s}_0) - {\boldsymbol i}),
    \label{eq:dirac_comb}
\end{equation}
where ${\boldsymbol s}_0$ specifies the relative position of the central pixel, and $\delta (\cdot)$ is the Dirac delta function. Throughout this paper, I work in units of native pixels; note that the Roman native pixel size is $0.11 \,{\rm arcsec}$. The distortion matrix ${\bf D}$ only captures linear terms of the geometric distortions of the focal plane, while world coordinate systems (WCSes) are usually written as fourth- or fifth-order polynomials. Nonetheless, according to dedicated tests during the development of {\sc Imcom} (see Footnote~38 of \papthree), such linear approximation is adequate within each $[{\cal O} (1) \,{\rm arcsec}]^2$ region of the sky, which is pertinent to the context of PSF reconstruction.

Real-world image devices are finite and have defects, hence only a finite subset of ${\mathbb Z}^2$ is sampled. This can be described as a window for ${\boldsymbol i}$ in Equation~(\ref{eq:dirac_comb}), which is omitted above for simplicity. Handling of missing pixels within the finite (usually square or rectangular) pixel array is discussed in Section~\ref{ss:tech}. For a given source position ${\boldsymbol r}$ and imaging device configuration (characterized by ${\boldsymbol s}_0$ and ${\bf D}$), the pixelated PSF $G' ({\boldsymbol i}; {\boldsymbol r})$ contains the same information as
\begin{equation}
    G' ({\boldsymbol s}; {\boldsymbol r}, {\boldsymbol s}_0, {\bf D}) = [G ({\boldsymbol s}; {\boldsymbol r}) * \Pi ({\boldsymbol s}; {\bf D})] \cdot \Sha ({\boldsymbol s}; {\boldsymbol s}_0, {\bf D}),
    \label{eq:pixelated}
\end{equation}
where $*$ denotes convolution. For simplicity, hereafter I omit the function parameters in the following discussion as long as the statement still makes clear intuitive sense; for instance, the right-hand side of Equation~(\ref{eq:pixelated}) simply reads $(G * \Pi) \cdot \Sha$.

The object $G * \Pi$ is worth some attention. On the one hand, it can be viewed as a lookup table for $G' ({\boldsymbol i}; {\boldsymbol r})$; on the other hand, it describes the probability distribution of the source direction of a photon landing at a given position. Both $G$ and $G * \Pi$ are instances of $({\mathbb R}^2, {\mathbb R}^2) \mapsto {\mathbb R}^+$; however, since such ``given position'' usually means the center of a pixel, $G * \Pi$ only needs to be defined as $({\mathbb R}^2, {\mathbb Z}^2) \mapsto {\mathbb R}^+$, where ${\mathbb R}^2$ is the space of source directions and ${\mathbb Z}^2$ is the space of pixel indices. I refer to $G$ (probability distribution of landing position given source direction) as a ``forward'' PSF and $G * \Pi$ (probability distribution of source direction given landing position) as a ``backward'' PSF. Such distinction is crucial for conceptualizing linear image regridding and coaddition.

Before concluding this section on the definitions of PSFs, here I comment on expected PSFs of the Roman Wide Field Instrument. The optical part has three major components: Airy disk due to the circular aperture, obscuration due to the secondary mirror, and diffraction spikes due to the struts supporting the secondary mirror. This part can be modeled using STPSF for Roman\footnote{\url{https://roman-docs.stsci.edu/simulation-tools-handbook-home/stpsf-for-roman}} developed at the Space Telescope Science Institute. In addition to the optical and pixelation parts, Roman PSFs also include effects of the H4RG-10 detectors \citep{2020JATIS...6d6001M}. These detector effects are expected to be largely reduced, if not eliminated, through calibration. For weak lensing cosmology purposes, PSFs will be measured from images of bright but unsaturated stars using software like PIFF \citep{2021MNRAS.501.1282J}. {\sc Imcom} \citep{2011ApJ...741...46R} takes PSFs in Roman images as input; specifically, it assumes that a PSF at a given position in a given image is known at high resolution a priori. This assumption is adopted throughout this work. Note that PSF reconstruction (see Section~\ref{ss:samp} below) does not add any new constraints on input PSF accuracy, although it may change the effect of having an inaccurate PSF model. Investigation into and mitigation of inaccuracies in PSF modeling are discussed in Section~\ref{ss:tech}.

\subsection{What are we undersampling?} \label{ss:samp}

Nyquist sampling (or beyond) is necessary for reliable shape measurements (see, e.g., Appendix~C of \papone). To determine whether an image is sufficiently sampled, the common practice is to compare its PSF width and pixel scale. The PSF width is usually characterized by $\xi \equiv \lambda / D$, where $\lambda$ is the wavelength of the observation and $D$ is the diameter of the entrance pupil. If the pixel scale is larger than half the PSF width in real space, it is narrower in Fourier space, and consequently some of the high-frequency Fourier modes cannot be unambiguously measured from the image, which is said to be ``undersampled.'' Otherwise, the image is ``oversampled,'' and in principle the information can be fully retrieved from it. Mechanically, it is possible to increase the spatial resolution (i.e., decrease the pixel scale) of an image via interpolation; however, such manipulation cannot increase the amount of information (see Section~\ref{ss:info} for further discussion). Furthermore, interpolating an undersampled image introduces discontinuities in the resulting (backward) PSF and hinders the accuracy of PSF-based measurements. As such, it is imperative that we make proper use of oversampled PSFs whenever available, as I explain in this section.

The finiteness of sampling is due to the discreteness and finite resolution of the pixel array.\footnote{Discreteness and finite resolution do not imply each other. A sampling can be discrete and have an infinite (i.e., arbitrarily high) resolution in the meantime; e.g., when the PSF $G$ in Equation~(\ref{eq:image_field}) is a Dirac delta function. Whether a set is discrete or not depends on the underlying continuum; for example, the set of all integers ${\mathbb Z}$ can be viewed as consecutive by itself.} Let us denote the true sky scene as $f ({\boldsymbol r})$; for a normalized point source in 2D, $f ({\boldsymbol r}) = \delta^2 ({\boldsymbol r})$. The image we obtain from an imaging device is
\begin{align}
    I = [f * G * \Pi + \eta] \cdot \Sha,
    \label{eq:image_field}
\end{align}
where $\eta$ is the noise field; in practice, an image is usually presented as signals in pixels:
\begin{align}
    I_{\boldsymbol i} \equiv [f * G * \Pi + \eta] ({\boldsymbol r}_{\boldsymbol i}).
    \label{eq:image_pixel}
\end{align}
Due to the $\Sha$ term in Equation~(\ref{eq:image_field}), information about $f$ encoded by $G * \Pi$ cannot be fully retrieved from the image $I$. However, oversampled PSFs are measured from a (usually large) collection of images and can be considered as external information while processing each individual image.

Linear image regridding and coaddition are formulated as a linear transformation from input signals $I_{\boldsymbol i}$ to an output signal $H_{\boldsymbol \alpha}$:
\begin{equation}
    H_{\boldsymbol \alpha} = \sum_{{\bar i}} \sum_{{\boldsymbol i} \in {\bar i}} T_{{\boldsymbol \alpha} {\boldsymbol i}}^{({\bar i})} I_{\boldsymbol i},
    \label{eq:transform}
\end{equation}
where ${\bar i} \subset {\mathbb Z}^2$ is the collection of available pixel indices in an input image, and $T_{{\boldsymbol \alpha} {\boldsymbol i}}^{({\bar i})}$ are the regridding/coaddition weights. Like ${\boldsymbol i}$, ${\boldsymbol \alpha} \in {\mathbb Z}^2$ is also written as a vector index in this work. Different algorithms differ because of different ways of determining these weights for selected pixels.

Recall from Section~\ref{ss:psf_def} that a backward PSF is defined as probability distribution of source direction given landing position. For each output pixel ${\boldsymbol \alpha}$, such a linear transformation also constructs an output (backward) PSF from properly shifted input (backward) PSFs:\footnote{In {\sc Imcom} papers, this is conventionally denoted as ${\rm PSF}_{\alpha, {\rm out}}$. Here I choose to introduce a new symbol for simplicity.}
\begin{equation}
    \Psi_{\boldsymbol \alpha} ({\boldsymbol s}) = \sum_{{\bar i}} \sum_{{\boldsymbol i} \in {\bar i}} T_{{\boldsymbol \alpha} {\boldsymbol i}}^{({\bar i})} G'_{\boldsymbol i} ({\boldsymbol r}_{\boldsymbol i} - {\boldsymbol R}_{\boldsymbol \alpha} + {\boldsymbol s}),
    \label{eq:recons_pixel}
\end{equation}
where ${\boldsymbol R}_{\boldsymbol \alpha}$ is the position of output pixel ${\boldsymbol \alpha}$, and $G'_{\boldsymbol i}$ is the backward PSF at the position of pixel ${\boldsymbol i}$ in image ${\bar i}$. To describe it in words, the meaning of Equation~(\ref{eq:recons_pixel}) is the probability distribution of source direction for the signal reallocated to output pixel ${\boldsymbol \alpha}$. Since Equation~(\ref{eq:transform}) is a sampling of an underlying $H$ field, the output pixelation function is just the Dirac delta function, and the output forward PSF is mathematically the same as the output backward PSF. Therefore, measurement algorithms naturally work on output images obtained in this way, as long as the output PSF is uniform across different output pixels, i.e.,
\begin{equation}
    \forall {\boldsymbol \alpha}: \Psi_{\boldsymbol \alpha} ({\boldsymbol s}) \simeq \Psi ({\boldsymbol s}),
    \label{eq:uniform}
\end{equation}
where $\Psi ({\boldsymbol s})$ is a unified function that does not depend on ${\boldsymbol \alpha}$.

To achieve the goal of PSF uniformity, {\sc Imcom} and Fast {\sc Imcom} allow the user to specify a target output PSF, $\Gamma ({\boldsymbol s})$, and optimize the weights $T_{{\boldsymbol \alpha} {\boldsymbol i}}^{({\bar i})}$ to minimize the discrepancy between $\Psi_{\boldsymbol \alpha}$ and $\Gamma$ for each output pixel ${\boldsymbol \alpha}$. Such discrepancy is quantitatively defined as the PSF leakage
\begin{equation}
    \frac{U_{\boldsymbol \alpha}}{C} \equiv \frac{\Vert \Psi_{\boldsymbol \alpha} - \Gamma \Vert^2}{\Vert \Gamma \Vert^2},
    \label{eq:leakage}
\end{equation}
where $\Vert \cdot \Vert$ is the $L^2$ norm. The discrepancy $\Psi_{\boldsymbol \alpha} - \Gamma$ in the numerator is referred to as the PSF residual; the denominator $C$ is introduced to make Equation~(\ref{eq:leakage}) dimensionless. {\sc Imcom} usually aims for $U_{\boldsymbol \alpha} / C \leq 10^{-6}$ (see Section~5.2 of \papone\ for the reason and some discussion). In \papfour, we found that a Gaussian target output PSF outperforms a smoothed Airy disk, and quality of the output images is better when the Gaussian is wide.

Since both target output PSF $\Gamma$ and (backward) input PSFs $G'_{\boldsymbol i}$ are oversampled, what is preventing us from obtaining zero leakage? To better understand this problem, let us make a reasonable approximation to Equation~(\ref{eq:recons_pixel}). The spatial variation of input PSFs is not drastic, hence it is a reasonable to assume that in the vicinity of a given position, $G'$ is the same for all pixels in each input image, i.e., $G'_{\boldsymbol i}$ can be simplified to $G'_{{\bar i}}$. Throughout \papone\ to \papfour, we only sampled $G' ({\boldsymbol s}; {\boldsymbol r})$ in ${\boldsymbol r}$ space once for each $2.5 \times 2.5 \,{\rm arcsec}^2$ region.\footnote{For readers familiar with {\sc Imcom}, such a region amounts to $2 \times 2$ ``postage stamps.''} Thus Equation~(\ref{eq:recons_pixel}) can be rewritten as
\begin{equation}
    \Psi_{\boldsymbol \alpha} \simeq \sum_{{\bar i}} (T^{({\bar i})} \cdot \Sha_{{\boldsymbol \alpha}}) * G'_{{\bar i}},
    \label{eq:recons_field}
\end{equation}
where $T^{({\bar i})}$ is some weight field for input image ${\bar i}$, and the subscript in $\Sha_{{\boldsymbol \alpha}}$ emphasizes that the Dirac comb is specific to each output pixel ${\boldsymbol \alpha}$. From Equation~(\ref{eq:recons_field}), it is clear that the limitation comes from the fact that we can only assign weights to a discrete set of positions, i.e., those of input pixels. In other words, we are potentially undersampling the underlying weight field.

To summarize, because of the discreteness and finite resolution of the imaging device, we are undersampling the convolution of true sky scene $f$ and pixelated PSFs $G * \Pi$. However, the PSFs, including both native PSFs and the user-specified target output PSF, are oversampled. The problem is that, when we perform linear image regridding or coaddition, some underlying weight field is also subject to finite sampling, limiting our ability to exactly reconstruct the desired PSF.

\subsection{Do we gain or lose information?} \label{ss:info}

In this section, I clarify two common concerns about potential changes in the amount of information due to image regridding and coaddition.

For a given survey, let us denote the total number of input pixels (in native images) as $N$ and the total number of output pixels (in regridded or coadded images) as $M$. While $N$ is determined by the instrument and survey design, in principle $M$ can be arbitrarily large; when $M > N$, there is an apparent gain of information. However, the linear transformation Equation~(\ref{eq:transform}) is a linear mapping ${\mathbb R}^N \mapsto {\mathbb R}^M$. Even if $M > N$, the image\footnote{In the linear algebra sense; not to be confused with an astronomical image.} of this mapping is at most $N$-dimensional, and the apparent gain is purely duplication. Due to the finite sampling in Equation~(\ref{eq:image_field}), different true sky scenes can lead to the same image(s). Image regridding and coaddition only change the representation of information and do not increase the amount of it.

That said, it is important to note that duplication does not mean fictitiousness. For example, when we use an appropriate transformation matrix to cast an image to a finer pixel grid, some higher-frequency modes (in Fourier space) become available. Since both regridding and Fourier transform are linear, each mode is still a linear combination of input signals and thus valid to some extent. This process is by no means ``creating'' fictitious information; instead, it is just combining information from the (potentially) undersampled input image and the oversampled PSF. In Equation~(\ref{eq:recons_field}), although $T^{({\bar i})} \cdot \Sha_{{\boldsymbol \alpha}}$ is discrete in real space and thus periodic in Fourier space, $G'_{{\bar i}}$ is continuous and non-periodic in both spaces. Therefore, their product in Fourier space, or equivalently their convolution in real space, is also continuous and non-periodic.

An opposite concern is that image regridding and coaddition may cause loss of information. From a linear algebra point of view, as long as output pixels outnumber input pixels ($M > N$) and the transformation matrices $T_{{\boldsymbol \alpha} {\boldsymbol i}}^{({\bar i})}$ are properly made, so that the dimension of the image of the linear mapping is $N$, such manipulations do not decrease the amount of information either. For PSF-fitting techniques like HSTPHOT \citep{2000PASP..112.1383D}, oversampled images with uniform and ``nicer'' (e.g., smoother, with monotonic radial profiles) PSFs may help avoid local minima of $\chi^2$ and improve the convergence rate.

Then what is an appropriate value of $M$ for a given $N$? On the one hand, $M$ needs to be larger than $N$ to preserve the amount of information; on the other hand, $M$ should not be arbitrarily large, as that would require a commensurate amount of storage, and extraneous duplication does not help measurements. Since the coverage (number of images overlapping with a given pointing) varies from position to position, the more relevant question is the number density, or equivalently size, of output pixels. Based on the above linear algebra argument, to avoid loss of information, we need to have
\begin{equation}
    s_{\rm out} \leq s_{\rm in} / (n_{\rm cover})^{1 / n_{\rm dim}},
    \label{eq:pixel_scale}
\end{equation}
where $s_{\rm out}$ and $s_{\rm in}$ are the output and input pixel scales, $n_{\rm cover}$ is the coverage ($\sim 6$ for most of HLIS), and $n_{\rm dim}$ is the spatial dimension ($2$ for real-world observations).

The situation is different for deep fields, i.e., fields with large $n_{\rm cover}$ values. For output images with pixel scale $s_{\rm out}$, the largest frequency of ``available'' Fourier modes seems to be $\upsilon_{\rm max} = 1 / (2 s_{\rm out})$; however, $\upsilon_{\rm max}$ is physically limited by the aperture size of the optical system. In other words, while deep fields can enhance both survey depth and spatial resolution, the latter has a physical upper limit, which should be taken into account while determining $s_{\rm out}$.

\subsection{1D PSFs used in this work} \label{ss:psf_1d}

To better illustrate ideas presented in Section~\ref{sec:essay} and specific algorithms to be introduced in Section~\ref{sec:meth}, this work mainly draws lessons from 1D. The 1D counterpart of a 2D Airy disk is a Fraunhofer single-slit diffraction pattern, and the corresponding PSF is
\begin{equation}
    G_{\rm 1D} (s) = \frac{[{\rm sinc}\, (s / \xi) - \varepsilon \,{\rm sinc}\, (\varepsilon s / \xi)]^2}{\xi (1 - \varepsilon)},
    \label{eq:single-slit}
\end{equation}
where $\xi \equiv \lambda / D$ is the ratio between wavelength of the observation and the diameter of the entrance pupil, and $\varepsilon$ is the linear obscuration. For Roman, $\varepsilon = 0.31$; for the H158 band, $\xi = 1.250$ native pixels. These values are adopted throughout this work. Following \papfour, the target output PSF form was chosen to be Gaussian
\begin{equation}
    \Gamma_{\rm 1D} (s) = \frac{e^{- s^2 / (2 \sigma^2)}}{\sqrt{2 \pi} \sigma},
    \label{eq:gaussian}
\end{equation}
where $\sigma$ is referred to as the ``width''; the corresponding full width at half maximum (FWHM) is simply $2 \sqrt{2 \ln 2} \sigma$. In \papfour, the benchmark width in the H158 band was $\sigma = 0.9343$ native pixels, and it was found that a larger width leads to more precise measurements. In most of this work, I use $\sigma = 1.8635$ native pixels, which is justified in Section~\ref{ss:width}. For simplicity and generality, the subscript ``1D'' is omitted below.

Given the (unpixelated) input PSF $G$ and the target output PSF $\Gamma$, the (ideal) weight field for image regridding $T$ should satisfy the equation
\begin{equation}
    \Gamma = G * \Pi * T,
    \label{eq:T_equation}
\end{equation}
which can be easily solved in Fourier space
\begin{equation}
    {\tilde T} = {\tilde \Gamma} / ({\tilde G} \cdot {\tilde \Pi}),
    \label{eq:T_solution}
\end{equation}
where ${\tilde \cdot}$ is Fourier transform. See Section~2.1 of \citet{2011ApJ...741...46R} for the {\sc Imcom} Fourier transform notation, which is adopted throughout this work. In practice, Equation~(\ref{eq:T_solution}) is computed based on discrete arrays (see Section~\ref{ss:setup} for the setup in this work) using fast Fourier transform (FFT). Since both Equations~(\ref{eq:single-slit}) and (\ref{eq:gaussian}) are symmetric (i.e., even functions), the imaginary parts of their Fourier transforms should be zero, and the non-zero values due to numerical errors are zeroed out. Furthermore, the integer-frequency modes of ${\tilde G} \cdot {\tilde \Pi}$ can be exactly zero, invalidating the division in Equation~(\ref{eq:T_solution}). Since ${\tilde \Gamma}$ is also a Gaussian function and decreases sharply at large $|x|$, high-frequency modes with $|\upsilon| > 1$ cycle per pixel are also zeroed out. Such zeroing-out operations have basically no impact on the resulting $T$ field.

\begin{figure}
    \centering
    \includegraphics[width=\columnwidth]{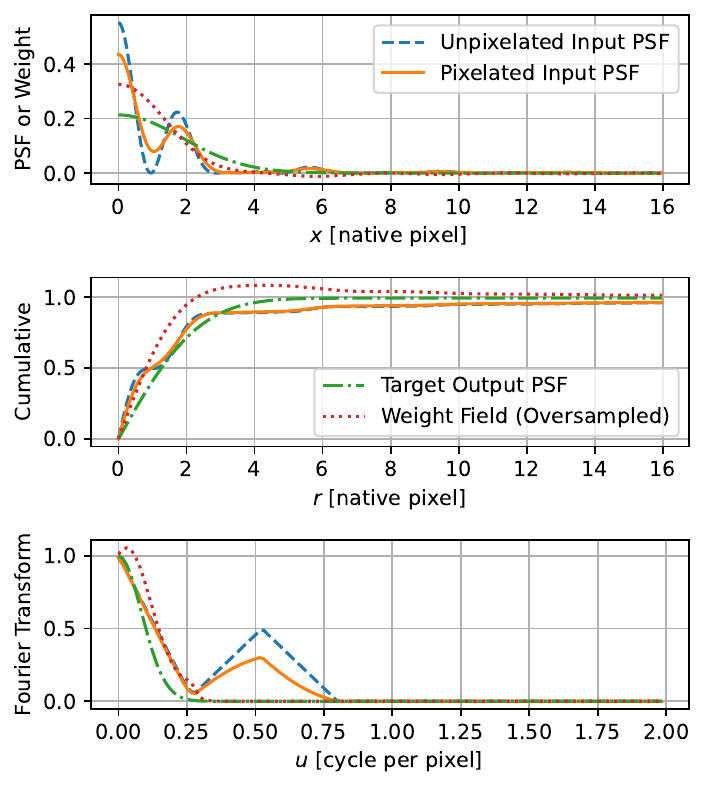}
    \caption{\label{fig:psfs_1d}1D PSFs used in this work. In each panel, the dashed blue (solid orange) curve represents the unpixelated (pixelated) input PSF, a obscured Fraunhofer single-slit diffraction pattern with $\xi \equiv \lambda / D = 1.250$ native pixels; the dash-dotted green curve represents the target output PSF, a Gaussian function with width $\sigma = 1.8685$ native pixels; the dotted red curve represents the oversampled weight field (see the text for explanation). The upper panel shows these four functions in real space; since they are all symmetric, only the $x \geq 0$ part is shown. The middle panel shows the total enclosed light (for PSFs) or weight as a function of the radius $r$; note that this includes both $x > 0$ and $x < 0$ parts. The lower panel shows the same functions in Fourier space; they are all purely real (i.e., no imaginary part), symmetric, and only non-zero at low frequencies.}
\end{figure} 

Figure~\ref{fig:psfs_1d} presents $G$ (blue dashed), $G * \Pi$ (orange solid), $\Gamma$ (green dash-dotted), and $T$ (red dotted) in three different ways. In real space (upper panel), we see that the first dark fringe of $G$ reaches zero but is smeared out in $G * \Pi$; the first bright fringe (the central one is considered the ``zeroth'') is more significant than usual due to linear obscuration. $\Gamma$ is wider than $G$ and $G * \Pi$ in terms of both FWHM (better seen in the upper panel) and half-light radius (better seen in the middle panel). In Fourier space (lower panel), it is clear that ${\tilde T}$ is the ratio between ${\tilde \Gamma}$ and ${\tilde G} \cdot {\tilde \Pi}$ because of Equation~(\ref{eq:T_solution}). Note that in 2D, the phenomenon of the modulation transfer function (magnitude of the Fourier transform of the PSF as a function of frequency) going down, then back up, and then down again is also possible, but it is not as pronounced as in the 1D case. At typical obscurations, it usually manifests as a ``shoulder'' rather than a ``trough'' like that at $u \sim 0.25$ cycle per pixel in Figure~\ref{fig:psfs_1d}. This difference might be important for some of the parameter choices.

The cumulative distributions (middle panel) contain noticeable information about the outer wings of PSFs. While the Gaussian $\Gamma$ quickly approaches (roughly defined as no visible discrepancy) unity, the single-slit $G$ and $G * \Pi$ do so in a much slower way. (Besides, the dark fringes of $G$ and $G * \Pi$ manifest as substantial inflection points in the cumulative distributions.) Even at $r = 32$ native pixels, the total enclosed light is only $98.87\%$ for both $G$ and $G * \Pi$. Similarly, Airy disks (in 2D) also have significant outer wings. This observation has several implications.
\begin{itemize}
    \item First, when we try to construct an oversampled image from a limited selection of input pixels, their total weights must be larger than $1$ to account for the fact that we are only using a portion of the input PSF. This explains why the total enclosed weight (shown in the middle panel) first increases and then decreases. In \papthree\ and \papfour, it was found that the ``total input weight'' ($\sum_{{\bar i}} \sum_{{\boldsymbol i} \in {\bar i}} T_{{\boldsymbol \alpha} {\boldsymbol i}}^{({\bar i})}$ using notation in this work) is typically larger than $1$ and is larger for the iterative kernel than for the Cholesky kernel. Now it is clear that this is because the ``acceptance radii'' (which characterize the spans of input pixel selection) are finite and are smaller for the iterative kernel.
    \item Second, this indicates that oversampled images with Gaussian PSFs are beneficial for finite-aperture\footnote{Here ``aperture'' means the size of image cutouts; not to be confused with the aperture of an instrument.} photometry. Using undersampled images with (pixelated) native PSFs, accounting for the aperture heavily relies on precise astrometry. This necessities multiple iterations, after which systematics due to undersampling may still remain.
    \item Third, for precision measurements, it is important to subtract outer wings, especially diffraction spikes, of bright objects before measuring its neighbors. See E. Macbeth et al. (2025, in preparation) for an {\sc Imcom}-based attempt using a novel ``PSF-splitting'' technique.
\end{itemize}

\section{Methodology} \label{sec:meth}

\begin{table*}[]
    \centering
    \caption{\label{tab:imcom_cf}Comparisons between {\sc Imcom} and Fast {\sc Imcom}. See referenced sections (including references therein) for details. The estimated computational costs and storage requirements are quoted for the Medium and Wide Tiers of the Roman HLIS \citep{2025arXiv250510574O}.}
    \begin{tabular}{llll}
    \hline
        Procedure or Functionality & {\sc Imcom} & Fast {\sc Imcom} & Reference(s) \\
    \hline
        Required inputs & Images with WCSes and PSFs & Same as {\sc Imcom} & Section~3 of \papone \\
        Uses oversampled PSFs to & Compute PSF correlations & Compute ideal weight fields & Sections~\ref{sec:essay} and \ref{sec:meth} \\
        Gets individual weights by & Building and solving linear systems & Sampling ideal weight fields & Sections~\ref{sec:essay} and \ref{sec:meth} \\
        Combines input images by & Directly assigning weights to pixels & Assigning meta-weights to images & Section~\ref{sec:meth} \\
        \multirow{2}{*}{Provided outputs} & \multirow{2}{*}{Images with {\sc Imcom} diagnostics} & \multirow{2}{*}{Same as {\sc Imcom}} & Section~4 of \papone\ and \\
        & & & Section~2.3 of \papthree \\
    \hline
        Est. computational costs & $\sim 100\,{\rm M}$ core-hours & $< 10\,{\rm M}$ core-hours & Section~\ref{ss:2d-fic} \\
        Est. storage requirements & $\sim 1.5 \,{\rm PB}$ (w/ accompanying images) & $\sim 0.2 \,{\rm PB}$ (Roman images only) & Section~\ref{ss:2d-fic} \\
        Survey optimization & Full simulations & Simple calculations & Section~\ref{ss:dither} \\
        Error propagation & Random realizations & Semi-analytic models & Section~\ref{ss:tech} \\
        Noise properties & Random realizations & Semi-analytic models & Section~\ref{ss:tech} \\
    \hline
    \end{tabular}
\end{table*}

In this section, I present two practical approaches to linear image regridding and coaddition. In Sections~\ref{ss:imcom} and \ref{ss:fast}, I explain the formalisms of {\sc Imcom} and Fast {\sc Imcom}, respectively. Table~\ref{tab:imcom_cf} compares some keys aspects of these two algorithms. Then in Section~\ref{ss:setup}, I describe the setup of the 1D experiments in this work.

Before diving into how these algorithms compute the weights $T_{{\boldsymbol \alpha} {\boldsymbol i}}^{({\bar i})}$ in Equations~(\ref{eq:transform}) and (\ref{eq:recons_pixel}), let me introduce another important diagnostic for regridded or coadded images. Although high-fidelity\footnote{In {\sc Imcom} papers, ``PSF fidelity'' is quantitatively defined as $- \log_{10} (U_{\boldsymbol \alpha} / C)$, where ``PSF leakage'' $U_{\boldsymbol \alpha} / C$ is defined in Equation~(\ref{eq:leakage}). This work directly addresses PSF leakage and does not use this quantitative definition, but still uses ``PSF fidelity'' to qualitatively mean ``good control over PSF leakage.''} PSF construction is desirable, this goal needs to be balanced with noise control. For each input image ${\bar i}$, the (input) noise covariance is denoted as $N_{{\boldsymbol i} {\boldsymbol j}}^{({\bar i})}$, and the output noise covariance is defined as
\begin{equation}
    \Sigma_{{\boldsymbol \alpha} {\boldsymbol \beta}}' = \sum_{{\bar i}} \sum_{{\boldsymbol i} \in {\bar i}} \sum_{{\boldsymbol j} \in {\bar i}} N_{{\boldsymbol i} {\boldsymbol j}}^{({\bar i})} T_{{\boldsymbol \alpha} {\boldsymbol i}}^{({\bar i})} T_{{\boldsymbol \beta} {\boldsymbol j}}^{({\bar i})};
    \label{eq:noise_cov}
\end{equation}
note that noise fields in different images are considered uncorrelated. To a good approximation, the noise field within each input image is uniform and uncorrelated, i.e., $N_{{\boldsymbol i} {\boldsymbol j}}^{({\bar i})}$ is an identity matrix multiplied by the (same) noise variance of every pixel. For an analysis of readout noise properties of Roman H4RG-10 detectors based on laboratory experiments, see \citet{2024PASP..136l4506L}. Furthermore, {\sc Imcom} optimizes weights on an output pixel-by-output pixel basis, and only the diagonal elements of Equation~(\ref{eq:noise_cov}) are directly minimized; the output noise correlation has been studied through simulated noise fields (see also \paptwo\ to \papfour\ for power spectra of coadded noise fields). Combining these considerations, the noise amplification is defined as
\begin{equation}
    \Sigma_{\boldsymbol \alpha} = \sum_{{\bar i}} \sum_{{\boldsymbol i} \in {\bar i}} \sum_{{\boldsymbol j} \in {\bar i}} T_{{\boldsymbol \alpha} {\boldsymbol i}}^{({\bar i})} T_{{\boldsymbol \alpha} {\boldsymbol j}}^{({\bar i})}.
    \label{eq:noise_amp}
\end{equation}
Despite its name,\footnote{Like that of the deceleration parameter $q$ in cosmology.} it is expected that image regridding/coaddition algorithms yield $\Sigma_{\boldsymbol \alpha} < 1$. This quantity then tells users to what extent output images are less noisy than input images and might have been better named ``noise deamplification.''

\subsection{{\sc Imcom} formalism} \label{ss:imcom}

{\sc Imcom} computes the $T_{{\boldsymbol \alpha} {\boldsymbol i}}^{({\bar i})}$ by building and solving linear systems. Following {\sc Imcom} convention, in this section I use scalar indices $i$ for input pixels and $\alpha$ for output pixels. These are flattened and concatenated (for input pixels only) version of the tuples $({\bar i}, {\boldsymbol i})$ and vector indices ${\boldsymbol \alpha}$, respectively.

For a given set of pixels, the {\sc Imcom} system matrices are defined and computed as
\begin{equation}
    A_{ij} = [G_j \otimes G_i] ({\boldsymbol r}_i - {\boldsymbol r}_j)
    \label{eq:imcom_Amat}
\end{equation}
and
\begin{equation}
    -\frac12 B_{\alpha i} = [\Gamma \otimes G_i] ({\boldsymbol r}_i - {\boldsymbol R}_\alpha),
    \label{eq:imcom_Bmat}
\end{equation}
respectively, where $\otimes$ denotes correlation. The solution is
\begin{equation}
    T_{\alpha i} = \sum_j [({\mathbf A} + \kappa_\alpha {\mathbf I})^{-1}]_{ij} \left( -\frac12 B_{\alpha j} \right),
    \label{eq:imcom_Tmat}
\end{equation}
where $\kappa_\alpha \geq 0$ is a (scalar) coefficient to balance the two minimization goals, PSF leakage Equation~(\ref{eq:leakage}) and noise amplification Equation~(\ref{eq:noise_amp}). As shown in the Appendix of \citet{2011ApJ...741...46R}, $U_\alpha$ monotonically increases with larger $\kappa_\alpha$, while $\Sigma_\alpha$ monotonically decreases. In \papthree, we found that it is reasonable to use a single value of $\kappa$ for all output pixels. This work prioritizes PSF reconstruction, hence I use $\kappa = 0$, which corresponds to minimum PSF leakage with {\sc Imcom}, throughout Sections~\ref{sec:regrid} and \ref{sec:coadd} and Appendix~\ref{app:asym}.

The {\sc Imcom} formalism supplies a shortcut of computing the PSF leakage:
\begin{equation}
    U_\alpha = \sum_{i, j} A_{ij} T_{\alpha i} T_{\alpha j} + \sum_i B_{\alpha i} T_{\alpha i} + C,
    \label{eq:imcom_Umat}
\end{equation}
where $C \equiv \Vert \Gamma \Vert^2$ is the square norm of the target output PSF. Because of the discreteness of the matrices, Equation~(\ref{eq:imcom_Umat}) is an approximation, and its reliability is also assessed in the results sections of this paper.

From Equations~(\ref{eq:imcom_Amat}) through (\ref{eq:imcom_Tmat}), one can see that {\sc Imcom} performs three time-consuming operations:
\begin{itemize}
    \item Fast Fourier transforms to compute the PSF correlations $G_j \otimes G_i$ and $\Gamma \otimes G_i$.
    \item Interpolations (see Appendix~A of \papone\ and Appendix B.1 of \papthree) to retrieve individual matrix elements $A_{ij}$ and $B_{\alpha i}$.
    \item Linear system solving (with Cholesky decomposition since \papthree) for obtaining the weights $T_{\alpha i}$.
\end{itemize}
The resulting computational complexity is discussed in Section~\ref{ss:2d-fic}, along with that of Fast {\sc Imcom}.

\subsection{Fast {\sc Imcom} formalism} \label{ss:fast}

According to Section~\ref{ss:imcom}, when there are multiple input images, {\sc Imcom} performs regridding (switching from the input pixel grid to the output pixel grid) and coaddition (combining all input images) at the same time. Fast {\sc Imcom} separates these two steps: It casts individual input images onto a common grid before combining them. In a sense, Fast {\sc Imcom} (like all other image regridding algorithms, including {\sc Imcom}) falls into the category of interpolation routines. Nevertheless, like {\sc Imcom}, Fast {\sc Imcom} explicitly has a target output PSF and the goal of noise control in ``mind'' while determining the interpolation weights.

In the first step,\footnote{When there is only one image to regrid, this is the only step.} Fast {\sc Imcom} constructs a regridded image from each input image:
\begin{equation}
    H_{\boldsymbol \alpha}^{({\bar i})} = \sum_{{\boldsymbol i} \in {\bar i}} T_{{\boldsymbol \alpha} {\boldsymbol i}}^{\prime ({\bar i})} I_{\boldsymbol i}.
    \label{eq:fast1_signal}
\end{equation}
This also constructs an intermediate PSF:\footnote{When there is only one image to regrid, this is the final PSF.}
\begin{equation}
    \Psi_{\boldsymbol \alpha}^{({\bar i})} ({\boldsymbol s}) = \sum_{{\boldsymbol i} \in {\bar i}} T_{{\boldsymbol \alpha} {\boldsymbol i}}^{\prime ({\bar i})} G'_{\boldsymbol i} ({\boldsymbol r}_{\boldsymbol i} - {\boldsymbol R}_{\boldsymbol \alpha} + {\boldsymbol s});
    \label{eq:fast1_pixel}
\end{equation}
using the $G'_{\boldsymbol i} \to G'_{{\bar i}}$ approximation (see Section~\ref{ss:samp}), this becomes
\begin{equation}
    \Psi_{\boldsymbol \alpha}^{({\bar i})} \simeq (T^{\prime ({\bar i})} \cdot \Sha_{{\boldsymbol \alpha}}) * G'_{{\bar i}},
    \label{eq:fast1_field}
\end{equation}
where the weight field $T^{\prime ({\bar i})}$ comes from Equations~(\ref{eq:T_equation}) and (\ref{eq:T_solution}). Naturally, the discrepancy $\Psi_{\boldsymbol \alpha}^{({\bar i})} - \Gamma$ is referred to as the intermediate PSF residual.

In the second step (of coaddition), a normalization factor, or ``meta-weight,'' is assigned to each intermediate image, so that the final signal is simply
\begin{equation}
    H_{\boldsymbol \alpha} = \sum_{{\bar i}} {\cal N}_{{\bar i}} H_{\boldsymbol \alpha}^{({\bar i})},
    \label{eq:fast2_signal}
\end{equation}
and similarly, the final output PSF is
\begin{equation}
    \Psi_{\boldsymbol \alpha} = \sum_{{\bar i}} {\cal N}_{{\bar i}} \Psi_{\boldsymbol \alpha}^{({\bar i})}.
    \label{eq:fast2_psf}
\end{equation}
It is expected that $\sum_{{\bar i}} {\cal N}_{{\bar i}} = 1$ because both $\Gamma$ and $\Psi_{\boldsymbol \alpha}^{({\bar i})}$ are normalized to unity, but in principle this sum can be slightly off.

Since there are two minimization goals, there is a spectrum of different strategies to determine these meta-weights. In this work, I study the two extreme strategies, as outlined below.
\begin{itemize}
    \item $U$-first strategy. Fast {\sc Imcom} can build and solve meta-linear systems for optimal meta-weights that minimize the final PSF leakage $U_{\boldsymbol \alpha}$. Intuitively, like Equation~(\ref{eq:imcom_Amat}), the meta-${\bf A}$ matrices should measure the correlations between intermediate PSFs, and like Equation~(\ref{eq:imcom_Bmat}), the meta-${\bf B}$ matrices should measure the correlations between intermediate PSFs and the target output PSF. However, the actual situation in this work is simpler, and specific forms of such meta-linear systems are given in Section~\ref{sec:coadd}.
    \item $\Sigma$-first strategy. Instead, Fast {\sc Imcom} can simply assign equal meta-weights\footnote{Roman enthusiasts should feel free to call this strategy ``equal-meta-{\bf w}eight-first,'' or ``w-first'' in short.} to all intermediate images, so that the noise amplification is minimized. Note that such equality assumes the intermediate noise amplification
    \begin{equation}
        \Sigma_{\boldsymbol \alpha}^{\prime ({\bar i})} = \sum_{{\boldsymbol i} \in {\bar i}} \sum_{{\boldsymbol j} \in {\bar i}} T_{{\boldsymbol \alpha} {\boldsymbol i}}^{\prime ({\bar i})} T_{{\boldsymbol \alpha} {\boldsymbol j}}^{\prime ({\bar i})}
        \label{eq:fast1_noise}
    \end{equation}
    is the same for all intermediate images. This assumption should be close to reality; otherwise, one can use the Lagrange multiplier method to minimize
    \begin{equation}
        \Sigma_{\boldsymbol \alpha} = \sum_{{\bar i}} {\cal N}_{{\bar i}}^2 \Sigma_{\boldsymbol \alpha}^{\prime ({\bar i})}.
        \label{eq:fast2_noise}
    \end{equation}
\end{itemize}
Both strategies are explored in Section~\ref{sec:coadd}, with nuances discussed therein. Since the $\Sigma$-first strategy is simple and seems robust, I think it will be favored under most practical circumstances (see Section~\ref{ss:tech} for discussion).

Non-extreme strategies balance these two goals. On the one hand, they have better control over noise than the $U$-first strategy but are not as good at PSF reconstruction. On the other hand, they have better control over PSF leakage than the $\Sigma$-first strategy but are not as good at noise control. This paper shows the extreme strategies to delineate the upper limits of PSF fidelity (achieved via the $U$-first strategy) and noise control (achieved via the $\Sigma$-first strategy), respectively, and thus inform the design of non-extreme strategies, which is left for future work.

\subsection{Setup in this work} \label{ss:setup}

For all 1D experiments in this work, an array of $64$ pixels is are ``excerpted'' from each input image. Note that $64 \,s_{\rm in} = 64 \times 0.11 \,{\rm arcsec} = 7.04 \,{\rm arcsec}$, while the largest span per dimension of pixel selection throughout \papone\ to \papfour\ was $3 \times 1.25 \,{\rm arcsec} = 3.75 \,{\rm arcsec}$.

The 1D PSFs have been introduced in Section~\ref{ss:psf_1d}. All input images are assumed to have the same PSF. The spatial resolution of the discrete representation is $1/32$, much finer than the $1/8$ adopted throughout \papone\ to \papfour. Note that large span and high resolution are only affordable for testing purposes.

Without loss of generality, I only study possible output pixel positions between the central two input pixels. Since I work in the units of native pixels, the relative positions of the $64$ input pixels are
\begin{equation}
    s = i + \Delta x, \quad i \in {\mathbb Z} \cap [-32, 31],
    \label{eq:setup}
\end{equation}
where $\Delta x \in [0, 1)$. When there are multiple images (i.e., in the case of coaddition), simple subscripts for input images are added to $\Delta x$, e.g., $\Delta x_0$. Note that Fast {\sc Imcom} treats every output pixel in the same way, excerpting input pixel arrays centering near its position. Therefore, this algorithm is not subject to postage stamp boundary effects, which result from the nonuniformity of input pixel windows (see \papthree). I revisit the topic of asymmetric windows in Appendix~\ref{app:asym} of this paper.

Furthermore, I only study $\Delta x$ values that are integer multiples of $1/32$, so that Equation~(\ref{eq:fast1_field}) can be computed without aliasing. Although that equation is written for Fast {\sc Imcom}, since weights are added to the same set of pixel positions, the actual reconstructed PSF of {\sc Imcom} can be computed using the same computer code.

\section{Image Regridding} \label{sec:regrid}

In this section, I explore topics involved in linear image regridding. In Section~\ref{ss:width}, I show how the width of the target output PSF affects the diagnostics of output images. In Section~\ref{ss:phase}, I study how the discrepancy between actual and target output PSFs changes as a function of the relative position of the output pixel. In Section~\ref{ss:rg-2d}, I provide evidence that results in 1D naturally extend to 2D.

\begin{figure*}
    \centering
    \includegraphics[width=\textwidth]{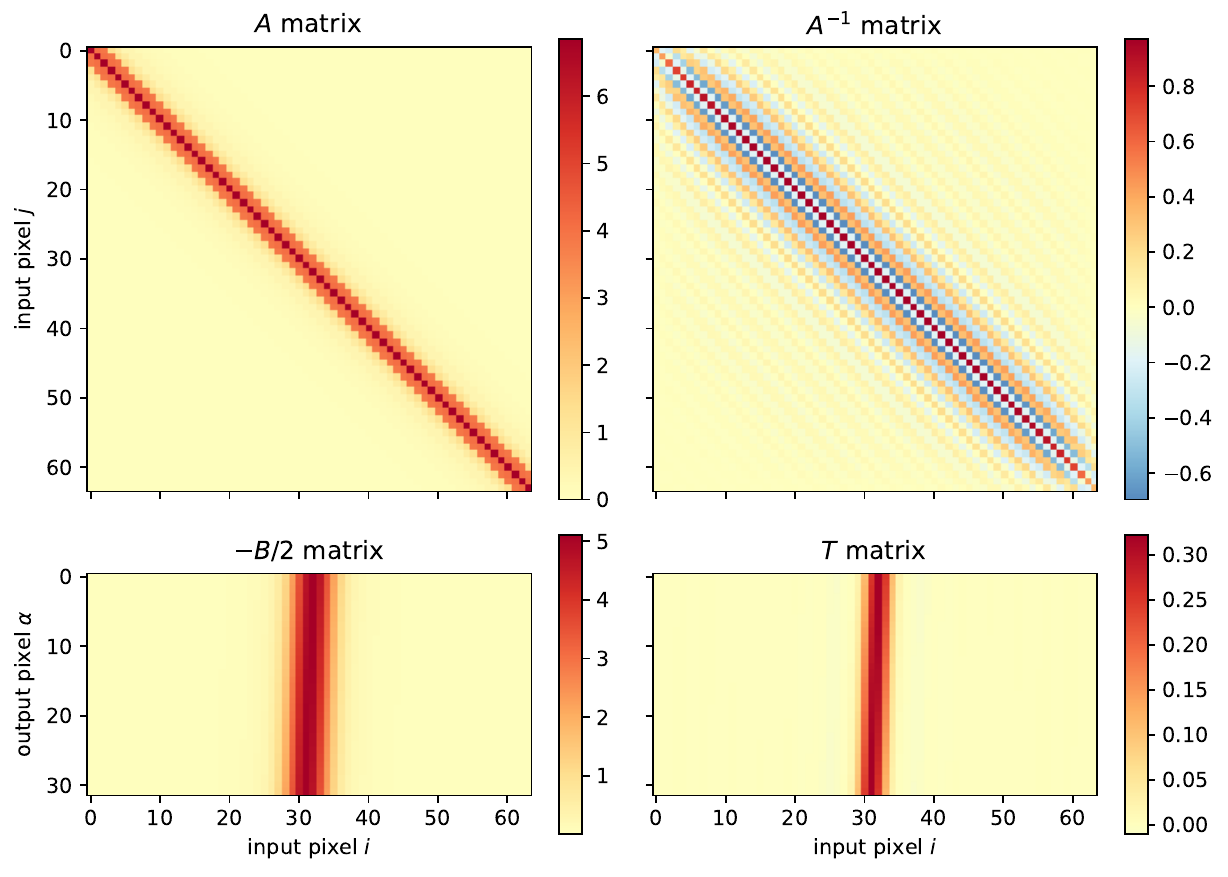}
    \caption{\label{fig:imcom_mats1}{\sc Imcom} matrices for image regridding. Upper row: the ${\bf A}$ matrix and its inverse ${\bf A}^{-1}$. Lower row: the $-{\bf B}/2$ matrix and the resulting ${\bf T}$ matrix.}
\end{figure*}

Figure~\ref{fig:imcom_mats1} presents a set of {\sc Imcom} matrices for image regridding. Compared to those in Figure~2 of \papthree, the ${\bf A}$ and $-{\bf B}/2$ matrices are much simpler for three reasons: first, the {\sc Imcom} matrices here are built for a 1D problem; second, only one input image is involved; third, the pixel array is not fragmented. All ${\bf A}$ matrix elements on a line parallel to the main diagonal (top left to bottom right) have the same value, and the antidiagonal (bottom left to top right) elements trace the autocorrelation of the pixelated input PSF $G * \Pi$. In this specific case, output pixel $\alpha$ corresponds to $\Delta x = \alpha / 32$, and each row of the $-{\bf B}/2$ matrix is a sampling of the cross correlation between the $G * \Pi$ and the target output PSF $\Gamma$. Note that in practice, the spacing between adjacent output pixels is almost always larger than $1/32$ of the input pixel scale. The ${\bf A}^{-1}$ matrix also manifests diagonal features, but with alternating positive and negative values in the antidiagonal direction. The resulting ${\bf T}$ matrix has such features in the horizontal direction (recall that each row is the weights of all input pixels for an output pixel), but the negative values have much smaller absolute values than positive ones, in agreement with the weight field shown in the upper panel of Figure~\ref{fig:psfs_1d}. Throughout Sections~\ref{ss:width} and \ref{ss:phase}, {\sc Imcom} results come from such matrices.

\subsection{Width of target output PSF} \label{ss:width}

\begin{figure*}
    \centering
    \includegraphics[width=\textwidth]{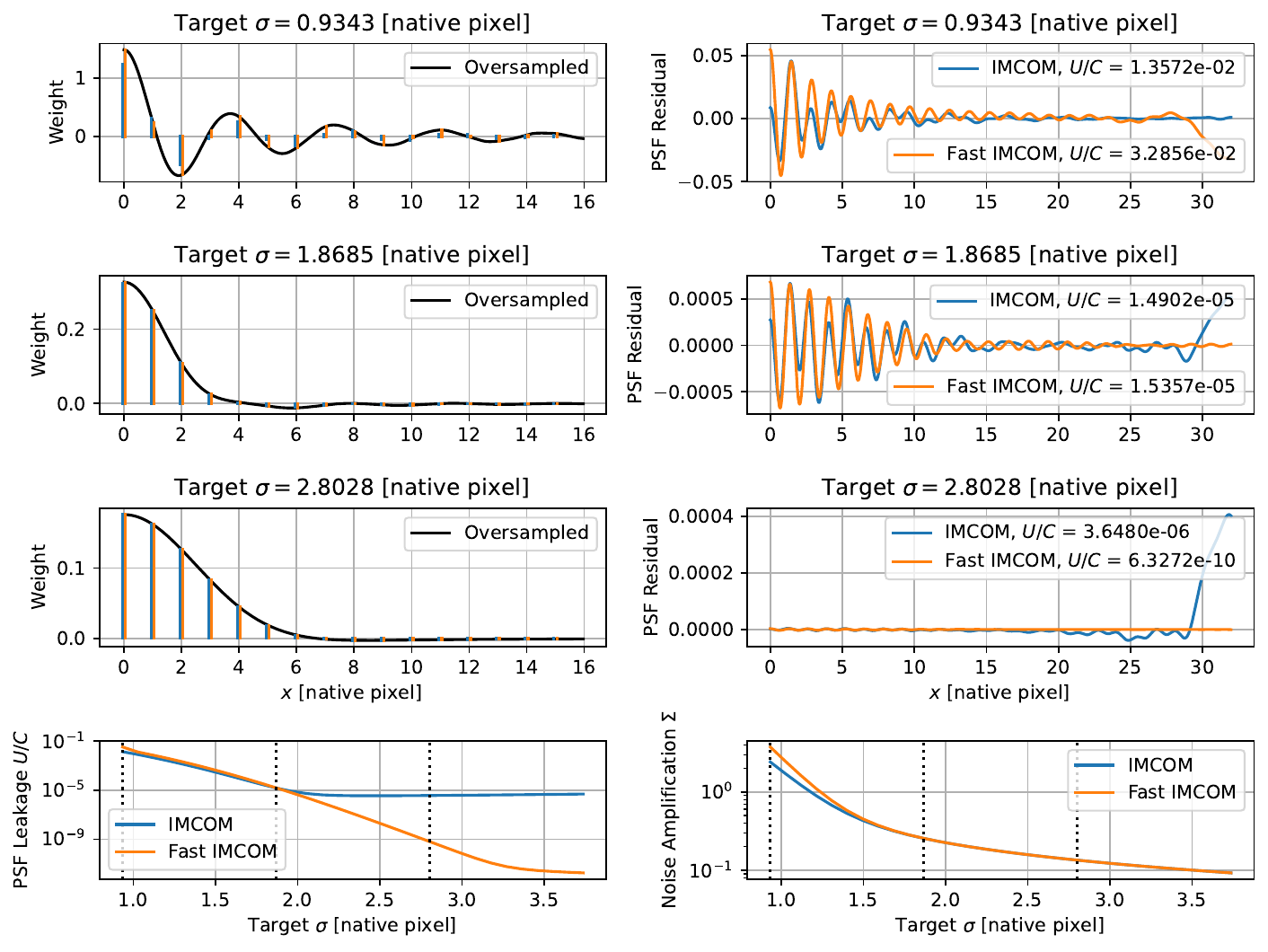}
    \caption{\label{fig:target_width}Impact of target output PSF width on image regridding. Each of the first three rows shows results for a different target width. The left panel includes the oversampled weight field (solid black curve) and the discrete weights determined by {\sc Imcom} (blue bars) and Fast {\sc Imcom} (orange bars); note that the discrete weights can only be added to the input pixel positions, but the bars are slightly displaced for clarity. The right panel presents the PSF residuals (reconstructed minus target) resulting from {\sc Imcom} (blue curve) and Fast {\sc Imcom} (orange curve) weights. This figure only includes results for an output pixel overlapping with one of the input pixels (i.e., $\Delta x = 0$), hence both weights and PSF residuals are symmetric, and only the $x \geq 0$ parts are shown. The last row shows the PSF leakage $U/C$ and noise amplification $\Sigma$ as a function of target output PSF width $\sigma$; the coloring is consistent with the preceding rows, and the three widths examined therein are marked with dotted black vertical lines.}
\end{figure*}

Figure~\ref{fig:target_width} explores the impact of target output PSF width on image regridding. The first three panels of the left column illustrate an idea discussed in Section~\ref{ss:samp}: By assigning weights to input pixels, we are sampling the underlying weight field (black curve). Fast {\sc Imcom} weights (orange bars) directly sample the weight field; {\sc Imcom} computes weights using Equation~(\ref{eq:imcom_Tmat}) and is not directly aware of the weight field, but the results are in good agreement with Fast {\sc Imcom}, especially at large $\sigma$. The first three panels of the right column display the corresponding PSF residuals; note that the PSF residuals are small compared to the target output PSF in all cases. From the first three rows, we see that the target output PSF width $\sigma$ determines the width of the weight field; since our sampling rate is always once per input pixel, larger $\sigma$ means that the weight field is better sampled. Therefore, the PSF leakage monotonically decreases with increasing $\sigma$, as shown in the bottom left panel. When $\sigma$ is small, the weight field has significant negative values; since the total weight is roughly the same (set by the size of the sampled portion of the input PSF), negative values lead to poor noise control because of Equation~(\ref{eq:noise_amp}). As $\sigma$ increases, the total weight is spread over more pixels, and the noise amplification decreases, as shown in the bottom right panel.

To summarize, a wider target output PSF leads to both better PSF fidelity and better noise control, in agreement with what we found in Section~5.1 of \papfour. Throughout the rest of this work, I adopt $\sigma = 1.8685$ native pixels, at which {\sc Imcom} and Fast {\sc Imcom} regridding results are of similar quality. With smaller $\sigma$, {\sc Imcom} performs slightly better in terms of both diagnostics, partially because {\sc Imcom} is directly minimizing the PSF leakage while Fast {\sc Imcom} is not. With larger $\sigma$, Fast {\sc Imcom} performs much better in terms of PSF reconstruction, while {\sc Imcom} soon encounters a barrier. This can be explained by Equation~(\ref{eq:imcom_Amat}): Although we have full information about the correlation $G_j \otimes G_i$, {\sc Imcom} samples it, which potentially causes loss of information. Both algorithms have the same control over noise at large $\sigma$.

\subsection{Relative position of output pixel} \label{ss:phase}

\begin{figure*}
    \centering
    \includegraphics[width=\textwidth]{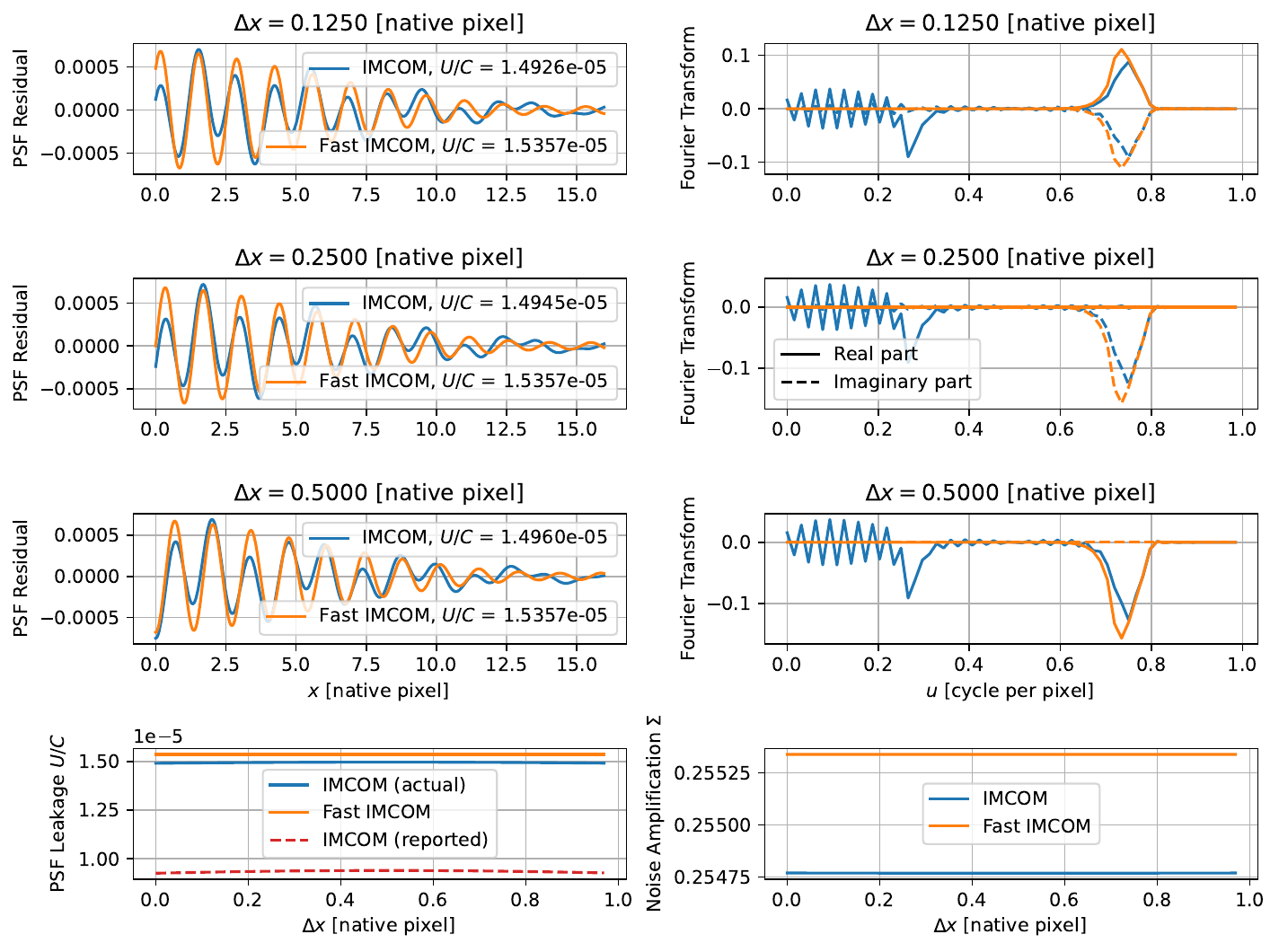}
    \caption{\label{fig:sub_shifts1}Impact of the relative position of the output pixel on image regridding. Each of the first three rows shows results for a different output pixel position, and the two panels show PSF residuals in real space (left) and Fourier space (right), respectively. Like in Figure~\ref{fig:target_width}, {\sc Imcom} results are shown in blue, while Fast {\sc Imcom} results are shown in orange. The last row shows the PSF leakage $U/C$ and noise amplification $\Sigma$ as a function of the relative position of the output pixel. For the former, the values reported by the {\sc Imcom} matrix formalism are shown as a dashed red curve (which is close to a horizontal line).}
\end{figure*}

Figure~\ref{fig:sub_shifts1} explores the impact of the relative position of the output pixel. From the first three panels of the left column, we see that Fast {\sc Imcom} PSF residuals are wave packets; with the same $\sigma$, the envelope of the wave packets is the same; at different $\Delta x$ values, their phases are different. In Fourier space, these wave packets have the same amplitudes but different complex phases. {\sc Imcom} PSF residuals and also wave packets; the phases are similar to those of the corresponding wave packets produced by Fast {\sc Imcom}, but they have an irregular envelope. In Fourier space, such irregularity manifests as low-frequency modes, which do not depend on $\Delta x$. The last row tells us that neither the PSF leakage nor the noise amplification depends on $\Delta x$; {\sc Imcom} slightly outperforms Fast {\sc Imcom}, but this is because of the adopted target output PSF width $\sigma$ (see Section~\ref{ss:width}).

Besides, the {\sc Imcom} approximation Equation~(\ref{eq:imcom_Umat}) slightly underestimates the PSF leakage, i.e., {\sc Imcom} ``thinks'' it is doing better than it actually is. As discussed at the end of Section~\ref{ss:width}, when {\sc Imcom} constructs the ${\bf A}$ matrix Equation~(\ref{eq:imcom_Amat}), there may be loss of information due to undersampling. However, Equation~(\ref{eq:imcom_Umat}) is based upon the (discrete) system matrices and does not ``know'' about the lost information. By this argument, Equation~(\ref{eq:imcom_Umat}) is an estimate of the lower limit of PSF leakage and needs to be treated with caution.

According to Figure~\ref{fig:sub_shifts1}, while Fast {\sc Imcom} does not seem as good as traditional {\sc Imcom}, it is actually more promising because of the simple pattern of its PSF residuals. In Fourier space, the residual is simply a constant profile multiplied by $e^{-2\pi i \Delta x}$ (here $i$ is the imaginary unit; note that $\Delta x$ is in units of native pixels). This indicates that:
\begin{itemize}
    \item For the coaddition of two images, if $|\Delta x_0 - \Delta x_1| = 1/2$, i.e., the two images are misaligned by exactly half a pixel, the residuals can exactly cancel out.
    \item For the coaddition of three non-overlapping images, even if none of the three possible $|\Delta x_i - \Delta x_j|$ values is exactly $1/2$, there is always a set of meta-weights to make the residuals exactly cancel out.
\end{itemize}
As for {\sc Imcom}, if we also write the final output image as a linear combination of intermediate images, the relatively high-frequency modes can be easily canceled out, but the low-frequency modes cannot. It turns out that these speculations agree with experimental results in Section~\ref{sec:coadd}.

\subsection{A glimpse at 2D} \label{ss:rg-2d}

\begin{figure*}
    \centering
    \includegraphics[width=\textwidth]{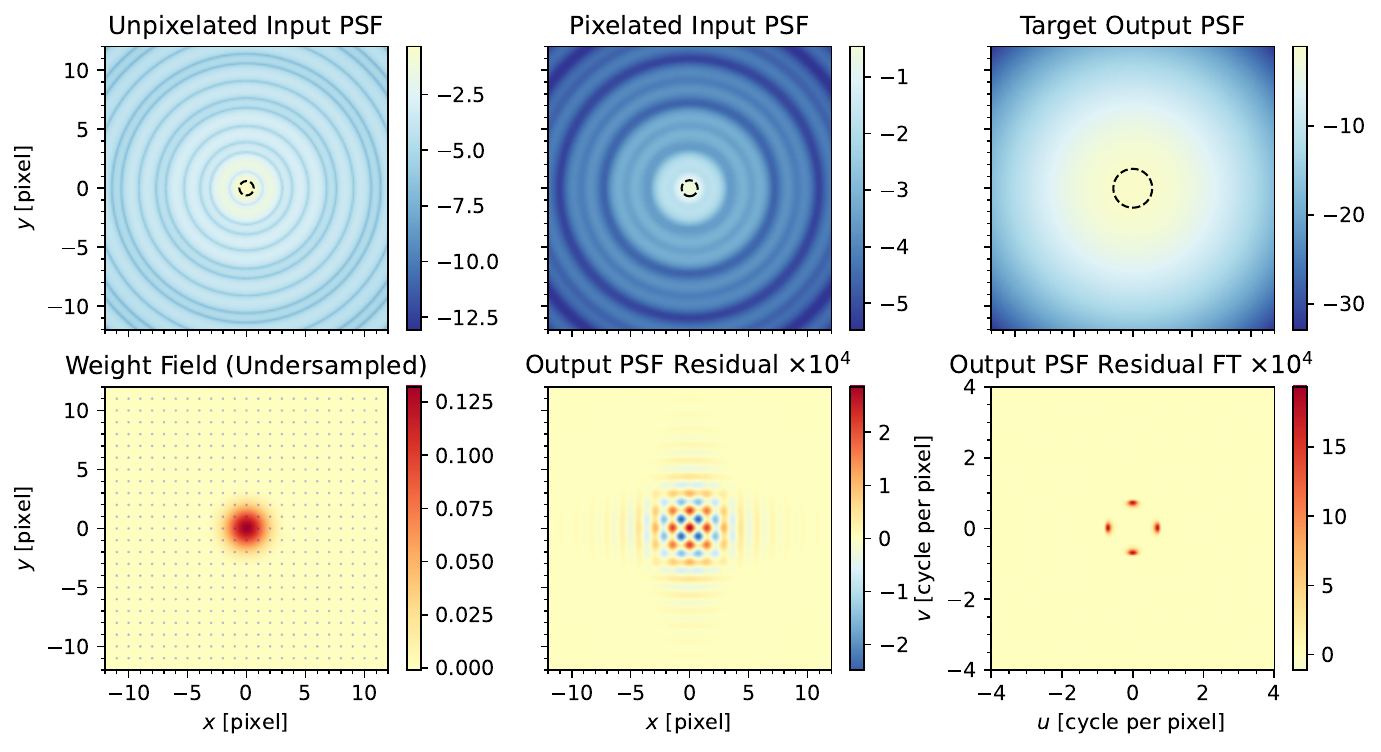}
    \caption{\label{fig:regrid_2d}PSFs, weight field, and PSF residual involved in 2D image regridding. The upper row presents the unpixelated (left) and pixelated (middle) versions of the input PSF, an obscured Airy disk with $\xi \equiv \lambda / D = 1.250$, as well as the target output PSF, a Gaussian function with width $\sigma = 1.4014$ native pixels. All three PSFs are shown in logarithmic scale, and their half widths at half maximum are shown as radii of dashed black circles. The lower left panel presents the oversampled weight field, along with the sampling points for the output pixel at $\Delta {\boldsymbol r} = (0, 0)^{\rm T}$ shown as blue points. The other two panels of the lower row show the resulting PSF residual in real space (middle) and Fourier space (right), respectively.}
\end{figure*}

While this paper mainly focuses on ``lessons from 1D,'' it is worth showcasing that Fast {\sc Imcom} in 2D is expected to produce simple PSF residual patterns as well. Figure~\ref{fig:regrid_2d} shows such an example; see Equation~(13) of \papfour\ for the general expression of a smoothed Airy disk, and note that Airy disks shown here are not smoothed and more closely resemble actual Roman PSFs in the H158 band. In real space (lower middle panel), the PSF residual is a 2D wave packet, as expected. The wave packet is not isotropic, because the 2D pixel grid has preferred directions, namely the $x$ and $y$ directions. In Fourier space (lower right panel), we see the 2D counterparts in $\pm x$ and $\pm y$ directions of the localized modes in Figure~\ref{ss:phase}. It is expected that the complex phases of the $\pm x$ and $\pm y$ modes are a constant profile multiplied by $e^{\mp 2\pi i \Delta x}$ and $e^{\mp 2\pi i \Delta y}$, respectively. The implications for dithering patterns are discussed in Section~\ref{ss:dither}, after I study image coaddition in 1D.

\section{Image Coaddition} \label{sec:coadd}

In this section, I study and discuss linear image coaddition. In Sections~\ref{ss:2-image} and \ref{ss:3-image}, I examine the coaddition of two and three 1D images, respectively. Then in Section~\ref{ss:2d-fic}, I describe how to generalize Fast {\sc Imcom} to 2D. In Section~\ref{ss:dither}, I explain the implications for dithering patterns of Roman surveys (extrapolated from 1D to 2D), which may apply to other instruments as well.

\begin{figure*}
    \centering
    \includegraphics[width=\textwidth]{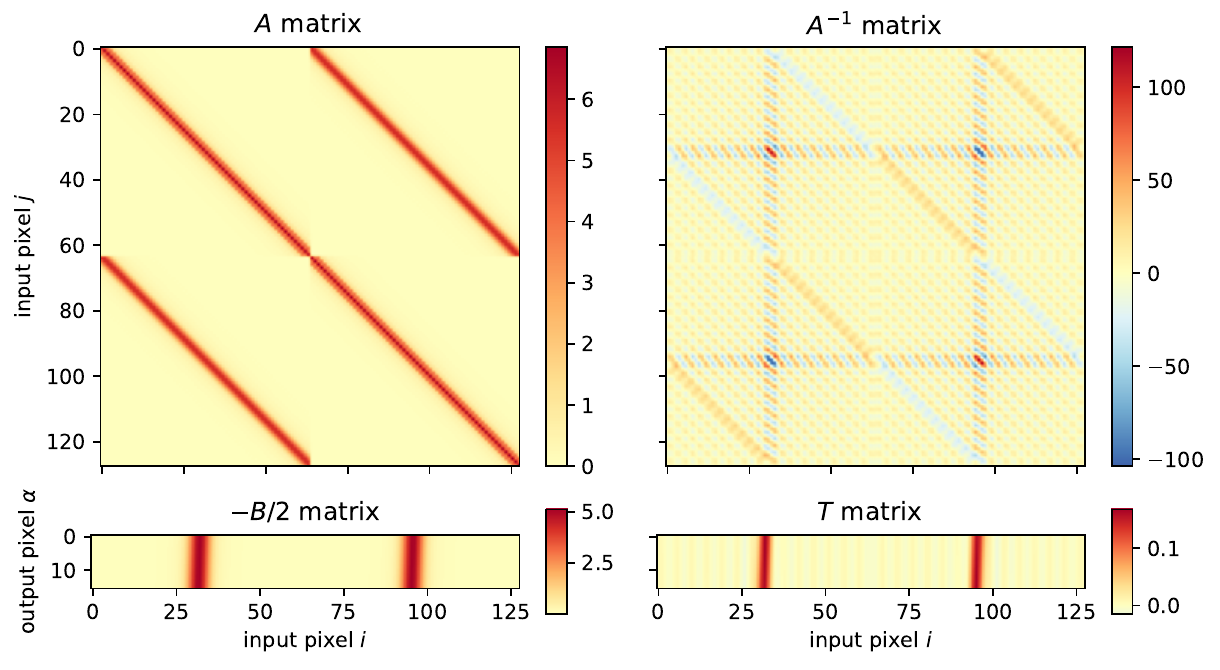}
    \caption{\label{fig:imcom_mats2}{\sc Imcom} matrices for the coaddition of two images, which are separated by $\Delta x_1 - \Delta x_0 = 1/2$. The layout is the same as that of Figure~\ref{fig:imcom_mats1}.}
\end{figure*}

Figure~\ref{fig:imcom_mats2} presents a set of {\sc Imcom} matrices for the coaddition of two images. The upper left and lower right quarters of the ${\bf A}$ matrix shown here are identical to that shown in Figure~\ref{fig:imcom_mats1}. Likewise, the left (right) half $-{\bf B}/2$ matrix shown here is identical to the upper (lower) half of that shown in Figure~\ref{fig:imcom_mats1}; the number of output pixels is halved to make sure that each of them is between the central two pixels of each input image, as described in Section~\ref{ss:setup}. As for the ${\bf A}^{-1}$ matrix, there are significant stripes in horizontal, vertical, and diagonal directions. While it is hard to develop an intuitive understanding of individual ${\bf A}^{-1}$ elements, we see that the resulting ${\bf T}$ matrix has similar patterns to those in Figure~\ref{fig:imcom_mats1}. Throughout Section~\ref{ss:2-image}, {\sc Imcom} results come from such matrices; the matrices for Section~\ref{ss:3-image} have different dimensions but do not contain new patterns, hence a dedicated figure is not included in this paper.

\subsection{Coaddition of two images} \label{ss:2-image}

\begin{figure*}
    \centering
    \includegraphics[width=\textwidth]{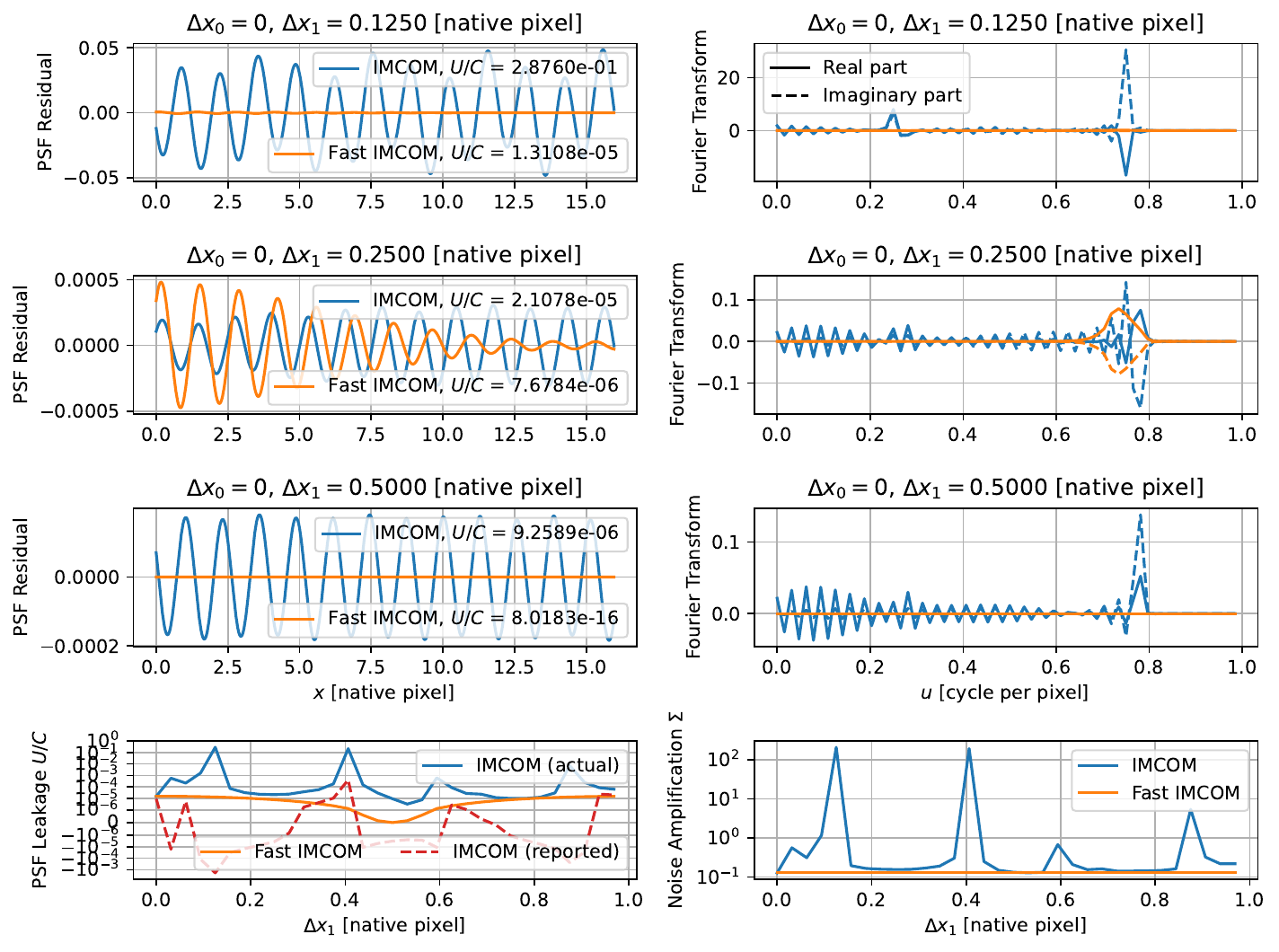}
    \caption{\label{fig:sub_shifts2}Impact of the relative positions of the input images on the coaddition of two images. The layout is similar to that of Figure~\ref{fig:sub_shifts1}; the difference is that the first three rows and the $x$-axes of the last row study the relative position of one of the input images ($\Delta x_1$) while keeping that of the other image ($\Delta x_0$) fixed.}
\end{figure*}

In the case of two images, the $U$-first and $\Sigma$-first strategies of Fast {\sc Imcom} (see Section~\ref{ss:fast}) give the same answer for the meta-weights: ${\cal N}_0 = {\cal N}_1 = 1/2$. Compared to the intermediate diagnostics while regridding each image, the PSF leakage is multiplied by $|\sum_{i=0}^1 e^{-2\pi i \Delta x_i}|^2 / 2^2 = \cos^2 (\pi |\Delta x_0 - \Delta x_1|)$, while the noise amplification is halved regardless of the separation $|\Delta x_0 - \Delta x_1|$. These theoretical expectations are met in Figure~\ref{fig:sub_shifts2}. The {\sc Imcom} results are less stable: For most possible separations, it performs almost as well as Fast {\sc Imcom}; for some ``unfortunate'' values like $0.1250$ native pixels, the results are severely corrupted by numerical instabilities. While a non-zero coefficient $\kappa_\alpha$ (see Section~\ref{ss:imcom}) would probably help, it is safe to conclude that Fast {\sc Imcom} is more robust and yields better results for the coaddition of two images. Besides, the {\sc Imcom} approximation Equation~(\ref{eq:imcom_Umat}) almost always fails to give the correct PSF leakage.

\subsection{Coaddition of three images} \label{ss:3-image}

In the case of three images, the two strategies of Fast {\sc Imcom} give different meta-weights. As mentioned in Section~\ref{ss:phase}, it is possible to get very-close-to-zero PSF leakage with the $U$-first strategy, which solves the meta-linear system
\begin{equation}
    \left\{ \begin{aligned}
    \sum_{i=0}^2 {\cal N}_i &= 1, \\
    \sum_{i=0}^2 {\cal N}_i \cos (2\pi \Delta x_i) &= 0, \\
    \sum_{i=0}^2 {\cal N}_i \sin (2\pi \Delta x_i) &= 0.
    \end{aligned} \right.
    \label{eq:fast_ufirst}
\end{equation}
Equation~(\ref{eq:fast_ufirst}) has exactly one solution when the three images are pairwise non-overlapping, i.e., when all three separations are non-zero. When $\Delta x_i = \Delta x_j \neq \Delta x_k$, where $\{ i, j, k \} = \{ 0, 1, 2 \}$, a reasonable set of meta-weights are ${\cal N}_i = {\cal N}_j = 1/4$ and ${\cal N}_k = 1/2$; the final PSF leakage is similar to that in the case of two images, while the noise control is better due to one additional image. When $\Delta x_i = \Delta x_j = \Delta x_k$, the only reasonable answer coincides with the universal answer of the $\Sigma$-first strategy: ${\cal N}_i = {\cal N}_j = {\cal N}_k = 1/3$. Compared to the intermediate diagnostics, the $\Sigma$-first PSF leakage is multiplied by $|\sum_{i=0}^2 e^{-2\pi i \Delta x_i}|^2 / 3^2$, and the noise amplification is divided by a factor of $3$.

\begin{figure*}
    \centering
    \includegraphics[width=\textwidth]{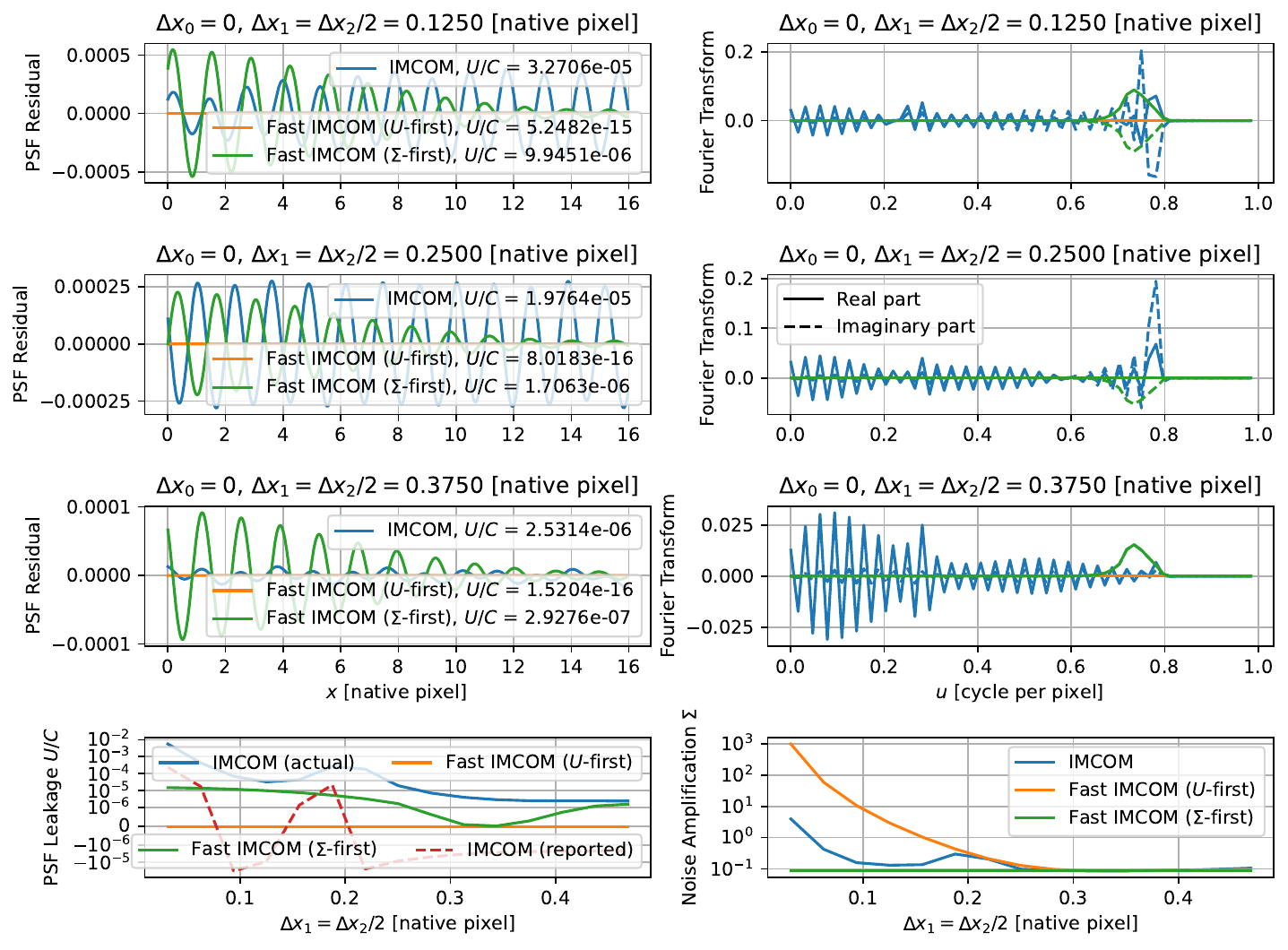}
    \caption{\label{fig:sub_shifts3}Impact of the relative positions of the input images on the coaddition of three images. The layout is similar to that of Figure~\ref{fig:sub_shifts1}; note that this figure only studies the special scenarios where the images are equally spaced ($\Delta x_0 = 0$, $\Delta x_2 = 2 \Delta x_1$). For the coaddition of three images, two extreme strategies of Fast {\sc Imcom}, $U$-first (prioritizing minimization of PSF leakage) and $\Sigma$-first (prioritizing minimization of noise amplification) are shown in orange and green, respectively.}
\end{figure*}

Figure~\ref{fig:sub_shifts3} explores a special category of configurations: $\Delta x_2 - \Delta x_0 = 2 (\Delta x_1 - \Delta x_0)$. Despite the particularity, the results manifest the behavior of each strategy. The $U$-first strategy of Fast {\sc Imcom} always yields a nearly perfect output PSF, as expected; however, the expense is that the noise amplification can be catastrophic when the separations are small. Intuitively, in this situation, Equation~(\ref{eq:fast_ufirst}) gives meta-weights with very large absolute values, with the middle image getting a negative meta-weight and the other two getting positive ones. The $\Sigma$-first strategy performs much better in terms of noise control, and the final PSF leakage is fully determined by the configuration of input images, as expected. {\sc Imcom} results are at most as good as those of the Fast {\sc Imcom} $\Sigma$-first strategy in terms of PSF fidelity. The situation of the approximation Equation~(\ref{eq:imcom_Umat}) is similar to the cases of image regridding (see Section~\ref{ss:phase}) and coaddition of two images (Section~\ref{ss:2-image}).

\begin{figure*}
    \centering
    \includegraphics[width=\textwidth]{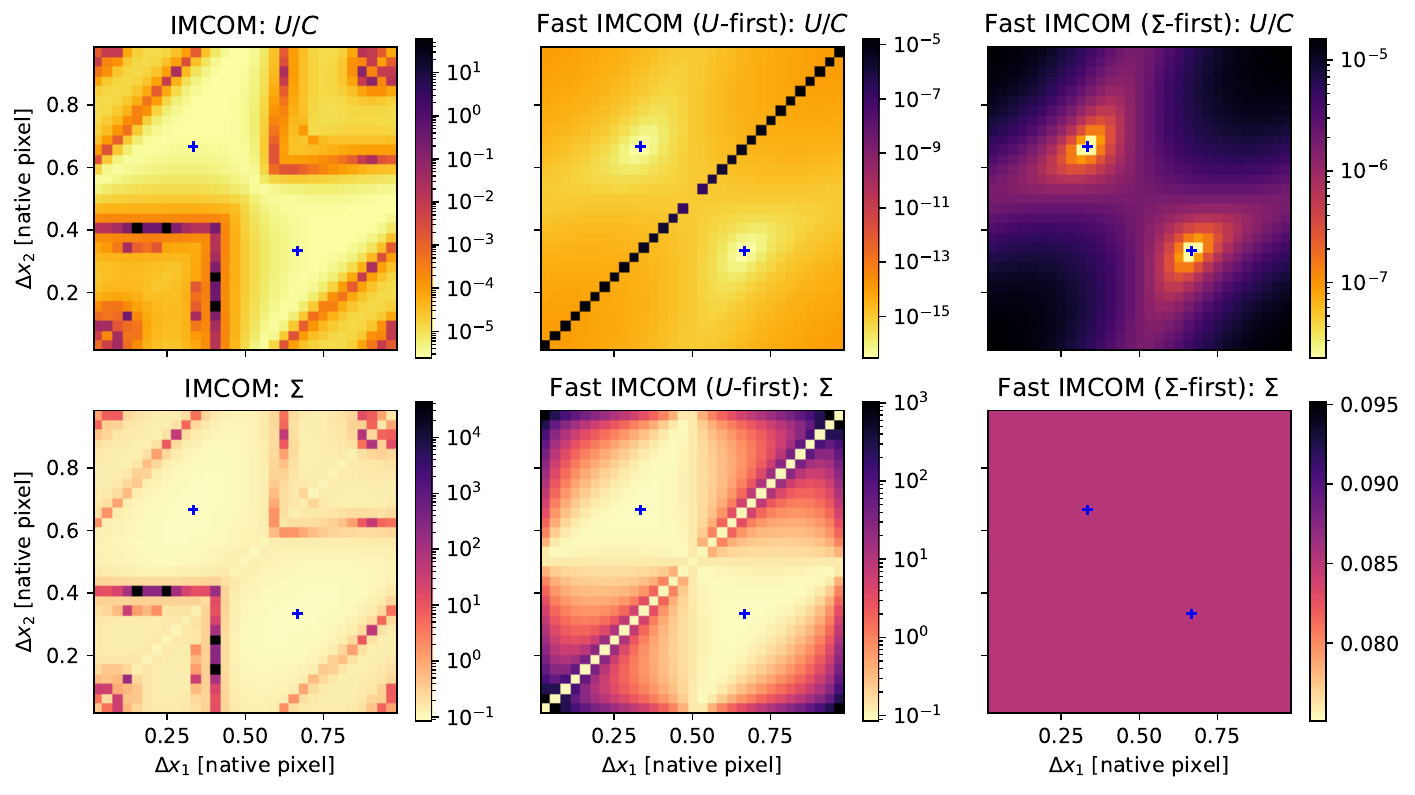}
    \caption{\label{fig:usigma_trade}An extended version of the last row of Figure~\ref{fig:sub_shifts3}. Each panel is map of the relative positions of two of the three input images ($\Delta x_1$ and $\Delta x_2$); that of the other image ($\Delta x_0 = 0$) is kept fixed. Specifically, the last row of Figure~\ref{fig:sub_shifts3} corresponds to the diagonal (from lower left to upper right) in this figure. The two rows presents PSF leakage $U/C$ and noise amplification $\Sigma$, respectively, and the three columns correspond to {\sc Imcom} and the two extreme strategies of Fast {\sc Imcom}. In each panel, the ideal dithering patterns for $\Delta x_0 = 0$, $\{\Delta x_1, \Delta x_2\} = \{1/3, 2/3\}$, are marked as blue plus signs.}
\end{figure*}

Figure~\ref{fig:usigma_trade} explores the full space of three-image configurations with $\Delta x_0 = 0$. It confirms our previous observations:
\begin{itemize}
    \item The $U$-first strategy of Fast {\sc Imcom} leads to negligible PSF leakage, except for the scenario where $\Delta x_1 = \Delta x_2$. Nevertheless, when any pair of images are close to each other (near the edges or near the diagonal), especially when all three images are close to each other (corners), the noise amplification is catastrophic.
    \item The $\Sigma$-first strategy of Fast {\sc Imcom} leads to a noise amplification value that does not depend on the configuration of input images, while the resulting PSF leakage varies from configuration (of separations) to configuration.
    \item In most cases, {\sc Imcom} results have reasonable PSF leakage and noise amplification. However, it is susceptible to numerical instabilities (again, the coefficient $\kappa_\alpha$ would probably help), and its results are basically never as good as those of the $\Sigma$-first strategy of Fast {\sc Imcom}.
\end{itemize}
It is noteworthy that with ideal configurations of input images ($\Delta x_0 = 0$, $\{\Delta x_1, \Delta x_2\} = \{1/3, 2/3\}$; blue plus signs), the $U$-first strategy and $\Sigma$-first strategies coincide again. In other words, the two minimization goals are no longer in contradiction with each other, and we can achieve nearly zero PSF leakage Equation~(\ref{eq:leakage}) and minimal noise amplification Equation~(\ref{eq:noise_amp}) at the same time. Therefore, results presented here motivate us to optimize the survey design. The extrapolated implications for 2D tiling patterns are discussed in Section~\ref{ss:dither}, after I describe how to extend Fast {\sc Imcom} to 2D.

With more than three pairwise non-overlapping images, the generalized version of Equation~(\ref{eq:fast_ufirst}) has an infinite number of solutions, and one can minimize the noise amplification while maintaining a near-zero PSF leakage. I stop at three images for this work.

\subsection{Extending Fast {\sc Imcom} to 2D} \label{ss:2d-fic}

While a 2D implementation of the Fast {\sc Imcom} algorithm is beyond the scope of this paper, here I describe some of its major aspects. Note that this section is a discussion based upon the findings in the previous sections; not everything is known a priori before seeing the 1D results.

The two-step procedure outlined in Section~\ref{ss:fast} is agnostic to the spatial dimensionality and thus applies to 2D as well. However, in addition to translational dithering, a 2D space also allows rotational dithering. Therefore, for each image, the mode groups in the lower right panel of Figure~\ref{fig:regrid_2d} are rotated by the roll angle of the image. Furthermore, in reality, the profile of the intermediate PSF residual is probably not a constant because of spatial variation of the input PSF and geometric distortions of the focal plane. Consequently, any counterpart to the meta-linear system Equation~(\ref{eq:fast_ufirst}) is just an approximation. To better handle the final PSF leakage, one needs to build a library of profiles of intermediate PSF leakages, and then the position-dependent complex phases are injected while building meta-linear systems for individual output pixels. Fortunately, according to the lower right panel of Figure~\ref{fig:regrid_2d}, the mode groups have finite sizes in both radial and angular directions. Therefore, when we coadd images, even if the input PSFs are slightly different, the intermediate PSF residuals can still largely cancel each other out. From a computational perspective, the library probably only needs to capture modes in a relatively thin annulus, so that the subsequent operations are not expected to be computationally expensive.

Since the algorithm is called Fast {\sc Imcom}, it is natural to ask how fast it is. Here I discuss the three time-consuming operations of {\sc Imcom} listed in Section~\ref{ss:imcom} one by one:
\begin{itemize}
    \item Fast Fourier transforms (FFTs): For $n_{\rm image}$ images covering the same area of the sky, {\sc Imcom} needs to perform ${\cal O} (n_{\rm image}^2)$ FFTs to compute $G_j \otimes G_i$ in Equation~(\ref{eq:imcom_Amat}). Fast {\sc Imcom} only needs to perform ${\cal O} (n_{\rm image})$ FFTs to compute Equation~(\ref{eq:T_solution}).
    \item Interpolations: For $n_{\rm pixel}$ selected input pixels and $m_{\rm pixel}$ planned output pixels, {\sc Imcom} needs to perform ${\cal O} (n_{\rm pixel}^2)$ interpolations to compute the ${\bf A}$ matrix using Equation~(\ref{eq:imcom_Amat}) and ${\cal O} (m_{\rm pixel} n_{\rm pixel})$ interpolations to compute the $-{\bf B}/2$ matrix using Equation~(\ref{eq:imcom_Bmat}). Fast {\sc Imcom} only needs to perform ${\cal O} (m_{\rm pixel} n_{\rm pixel})$ interpolations to directly obtaining the $T$ matrix, i.e., the sampled weight fields in Equation~(\ref{eq:fast1_field}). Furthermore, Fast {\sc Imcom} interpolations are expected to benefit more from the regularity of the pixel array.
    \item Linear system solving: With the Cholesky kernel of {\sc Imcom}, the complexity of decomposing an ${\bf A}$ matrix is ${\cal O} (n_{\rm pixel}^3 / 6)$, while that of applying the decomposed version to output pixels is ${\cal O} (m_{\rm pixel} n_{\rm pixel}^2)$. As for Fast {\sc Imcom}, the size of meta-linear systems is $n_{\rm image} \times n_{\rm image}$; since $n_{\rm pixel} = {\cal O} (10^3) \,n_{\rm image}$, the time consumption of solving them is negligible. That said, depending how PSF leakages are tracked, the time consumption of building them can be larger.
\end{itemize}
Besides, both algorithms need to perform other operations. In conclusion, a precise answer to the question ``how fast is Fast {\sc Imcom}'' is not available without tests. An educated speculation is that Fast {\sc Imcom} can be about an order of magnitude faster than {\sc Imcom}.

The Roman HLIS will cover $2415 \,{\rm deg}^2$ of the sky in three bands (Medium Tier) and additional $2702 \,{\rm deg}^2$ in the H158 band (Wide Tier) \citep{2025arXiv250510574O}. The total computational costs of processing all these images with {\sc Imcom} are currently estimated to be $\sim 100\,{\rm M}$ core-hours. Since this is expensive, the current plan is to store multiple (${\cal O} (10)$) versions of simulated objects and noise fields along with actual coadded Roman images, and the total storage requirements are $\sim 1.5 \,{\rm PB}$. If Fast {\sc Imcom} is to be used instead of {\sc Imcom}, these accompanying images (``layers'' in the {\sc Imcom} terminology) can be produced during analysis and do not need to be permanently stored. Hence Fast {\sc Imcom} has the potential of reducing the storage requirements by an order of magnitude (to $\sim 0.2 \,{\rm PB}$) as well.

\subsection{Implications for dithering patterns} \label{ss:dither}

According to Section~\ref{ss:3-image}, the $\Sigma$-first strategy of Fast {\sc Imcom} is more robust than the $U$-first strategy. If the former is to be adopted, a good control over noise is guaranteed, while the PSF fidelity is determined by the dithering pattern. Therefore, the implications of findings in this work for dithering patterns (extrapolated from 1D to 2D) are worth discussing.

Here I define some terms to describe the relationship between a set of images.
\begin{itemize}
    \item If a set of images share the same $x$- and $y$-axes, they are {\bf coherent}. Note that in the lower right panel of Figure~\ref{fig:regrid_2d}, the mode groups have a finite angular size, hence even if the roll angles of two images are slightly different, they can still be coherent. According to the HLIS dithering pattern \citep[see, e.g., Appendix~A of][]{2023MNRAS.522.2801T}, this is likely the case for a group of gap-filling dithers. Otherwise, when a set of pairwise incoherent images are combined, equal meta-weights should be assigned, and the PSF leakage is inversely proportional to the number of input images.
    \item If a pair of images are coherent and separated by the same fraction (complementary fractions) of a pixel in the two directions, they are {\bf phase (anti-)locked}. In general, the 2D counterpart of Equation~(\ref{eq:fast_ufirst}) contains five equations (including two additional ones for $\Delta y_i$), but a group of translational dithers usually does not have five members. If a group of coherent images are pairwise phase locked, the number of equations is reduced back to three, and three such images allow for a very-close-to-zero PSF leakage, like the $U$-first results in Section~\ref{ss:3-image}.
    \item If a pair of images are coherent and separated by half a pixel in both directions (one direction), they are in {\bf double (single) resonance}. For two images, double resonance is ideal, as equal meta-weights allow for almost perfect PSF reconstruction from two images, like in the third row of Figure~\ref{ss:phase}. Such a pattern is particularly desirable for the Roman Galactic Plane Survey, in which there are only two images in each band \citep{2025arXiv251107494G}.
\end{itemize}

While these patterns are advantageous, it is difficult to uniformly secure any of them throughout the entire focal plane due to geometric distortions. Optimization of dithering patterns of specific surveys is left for future work.

\section{Discussion} \label{sec:disc}

In this section, I further discuss topics related to Fast {\sc Imcom}. Technical issues and scientific applications are addressed in Sections~\ref{ss:tech} and \ref{ss:sci}, respectively.

\subsection{Technical discussion} \label{ss:tech}

\paragraph{Handling missing pixels} Throughout this work, I have been assuming that the input pixel arrays are complete. Yet in reality, an imaging device always have some ``inoperable'' pixels. For example, the Roman Wide Field Instrument contains $\sim 3\%$ permanent bad pixels (see Table~2 of \papone); furthermore, a pixel can be temporarily unreliable due to cosmic ray hits or persistence. {\sc Imcom} treats input images as individual pixels, and missing pixels simply amount to a reduction in the dimensionality of linear systems. Fast {\sc Imcom} treats input images as arrays of pixels, hence missing pixels are worth some more attention.

Let us consider a pixel array with only one missing pixel. During regridding, for a given output pixel ${\boldsymbol \alpha}$, if the missing pixel is suppose to carry weight $T_{{\boldsymbol \alpha} {\boldsymbol i}}^{({\bar i})}$, setting the weight to zero causes a missing addend in Equation~(\ref{eq:fast1_pixel}). This increases the intermediate PSF leakage by about $(T_{{\boldsymbol \alpha} {\boldsymbol i}}^{\prime ({\bar i})})^2 \Vert G'_{\boldsymbol i} \Vert^2 / \Vert \Gamma \Vert^2$. Similarly, when there are multiple missing pixels, the total increase of the intermediate PSF leakage is roughly proportional to the sum of the squares of the weights that the missing pixels are supposed to carry. In practice, a reasonable strategy is to set a threshold for such a sum, which needs to be optimized via tests for each target output PSF. As seen from Figure~6 of \papthree\ and Figure~4 of \papfour, when missing pixels are supposed to carry significant weights, the corresponding image is not able to contribute much to the reconstruction anyway.

Furthermore, since Fast {\sc Imcom} is expected to be fast (see Section~\ref{ss:2d-fic}), it should be relatively inexpensive to execute iterative schemes. Specifically, it is possible to ``fix'' missing pixels using previous coaddition results. While this scheme allows us to make better use of existing pixels, it also makes the noise covariance hard to track. Therefore, different science purposes may favor different choices.

\paragraph{{\sc Imcom} to-do list} In Section~6 of \paptwo, we included a list of {\sc Imcom} items that need to be studied. Recently, in Section~6 of \papfour, we made some updates to the list. In addition to enhancing computational efficiency (first item) and making deep fields (seventh item) easier to handle, the advent of Fast {\sc Imcom} has the potential of facilitating several other research projects.
\begin{itemize}
    \item Error propagation (third item): This includes ``propagation of astrometric errors, relative flux calibration between images, and PSF model errors.'' In Equation~(\ref{eq:fast1_pixel}), astrometric errors are errors in ${\boldsymbol r}_{\boldsymbol i}$, flux calibration amounts to the overall scaling of $\Psi_{\boldsymbol \alpha}^{({\bar i})}$, and PSF model errors affect $T_{{\boldsymbol \alpha} {\boldsymbol i}}^{\prime ({\bar i})}$ via Equation~(\ref{eq:T_solution}). Therefore, the mathematical framework of Fast {\sc Imcom} allows for semi-analytic investigations of these issues. Furthermore, the high efficiency of Fast {\sc Imcom} may allow us to establish a feedback loop to iteratively correct these errors.
    \item Noise fields (fourth item): While \citet{2024PASP..136l4506L} and \paptwo\ to \papfour\ all addressed noise properties in coadded images, noise power spectra were only computed for each block (with side length at the ${\cal O} (1) \,{\rm arcmin}$ level; see Table~1 of \papfour\ for a summary). For high-precision measurements, the specific noise covariance in the vicinity of an object may be needed. According to the Fast {\sc Imcom} formalism, the noise covariance matrix in a regridded image can be retrieved (via interpolation) from the autocorrelation of the weight field, and that in coadded images is a linear combination of intermediate covariance matrices.
    \item Chromatic effects (sixth item): See \citet{2025MNRAS.542..608B} for a study on chromaticity in the context of Roman weak gravitational lensing cosmology. Throughout \papone\ to \papfour, we assumed flat spectral energy distributions (SEDs) while making input PSFs. Running {\sc Imcom} multiple times with different SEDs is a useful way of characterizing and potentially mitigating the impact of chromatic PSFs, yet the computational costs of running {\sc Imcom} once are already tremendous. Fast {\sc Imcom} may change this scenario (see Section~\ref{ss:2d-fic}).
\end{itemize}

\paragraph{Fast {\sc Imcom} strategy} Like {\sc Imcom}, Fast {\sc Imcom} has two minimization goals, PSF leakage Equation~(\ref{eq:leakage}) and noise amplification Equation (\ref{eq:noise_amp}). There is a spectrum of strategies to assign meta-weights to images (see Section~\ref{ss:fast}), of which I have investigated two extremes, the $U$-first strategy, which minimizes PSF leakage, and the $\Sigma$-first strategy, which minimizes noise amplification. Based upon results in Section~\ref{sec:coadd} and above discussions in this section, here I argue that the latter should be favored for the following reasons.
\begin{itemize}
    \item Robustness. As shown in Section~\ref{ss:3-image}, in particular Figure~\ref{fig:usigma_trade}, for unfavorable configurations of input images, PSF leakage values yielded by the $\Sigma$-first strategy are still under control, while noise amplification values yielded by the $U$-first strategy are sometimes catastrophic. The latter is unavoidable as long as we try to minimize PSF leakage at the image level.
    \item Efficiency. As discussed in Section~\ref{ss:2d-fic}, linear system solving is the third potentially time-consuming operation. However, since the $\Sigma$-first strategy tends to assign equal meta-weights to available input images, it does not involve linear system solving and brings a speedup compared to the $U$-first strategy or any other strategy in between.
    \item Simplicity. To enable semi-analytic studies of error propagation and noise properties (see above) or survey optimization (see Section~\ref{ss:dither}), meta-weights need to be tracked. If we assign equal meta-weights to input images, we only need to track whether an input image is used for an output pixel (Boolean arrays that support lossless compression); otherwise, we would have to record all specific meta-weights (floating-point number arrays that can only be compressed in lossy ways). Therefore, the $\Sigma$-first strategy is strongly favored from a storage perspective.
\end{itemize}
Whether the $\Sigma$-first strategy is sufficiently good in practice will be investigated in a follow-up work (K. Cao et al. 2026, in preparation).

\subsection{Scientific discussion} \label{ss:sci}

\paragraph{Other weak lensing programs} In addition to Roman, major facilities for Stage IV weak lensing programs also include the Legacy Survey of Space and Time (LSST) at the NSF-DOE Vera C. Rubin Observatory \citep{2012arXiv1211.0310L, 2019ApJ...873..111I} and the Euclid space telescope \citep{2011arXiv1110.3193L, 2022A&A...662A.112E, 2024arXiv240513491E}. Here I briefly discuss the potential applications of Fast {\sc Imcom} to both programs.

LSST is conducted with a ground-based instrument, hence its PSFs are largely determined by seeing conditions of the Earth's atmosphere at the time of observation. Thanks to a dedicated auxiliary telescope, LSST is expected to have a good PSF model. It is potentially beneficial to coadd LSST images with {\sc Imcom} to obtain a uniform PSF with a simple form, yet the LSST coverage can be two to three orders of magnitudes larger than that of Roman, so it is prohibitively expensive. Fast {\sc Imcom} is expected to be much faster, especially when $n_{\rm image}$ is large (see Section~\ref{ss:2d-fic}), and worth trying. Because of such large coverage, the $\Sigma$-first strategy of Fast {\sc Imcom} should be favored. On the one hand, when the distribution of roll angles is close to randomness (or isotropy), the situation is similar to the ``pairwise incoherent'' scenario mentioned in Section~\ref{ss:dither}, and the PSF leakage is inversely proportional to the number of input images, which is usually good enough given the large number. On the other hand, the $U$-first strategy or any other strategy in between would require building and solving meta-linear systems with size length at the ${\cal O} (10^3)$ level for each output pixel, which could be computationally expensive.

Euclid is also a space mission and thus has stable PSFs. For weak lensing purposes, one of its main limitations comes from chromaticity due to its wide filters. As discussed in Section~\ref{ss:tech}, Fast {\sc Imcom} may provide a reasonable solution to chromatic PSFs and thus enhance the precision of archival data analysis. The Euclid Wide Survey (WS) has a typical coverage of three or four and the dithers are purely translational. In the terminology defined in Section~\ref{ss:dither}, the set of three or four images are supposed to be pairwise coherent. In this case, the $\Sigma$-first strategy of Fast {\sc Imcom} may cause relatively large PSF leakage (see the upper right panel of Figure~\ref{fig:usigma_trade}), and the $U$-first strategy or another strategy in between is probably favored.

\paragraph{Roman time domain surveys} Roman will implement three Core Community Surveys. In addition to the High Latitude Wide Area Survey, of which the High Latitude Imaging Survey is a component, there are two time domain surveys, namely the High Latitude Time Domain Survey (mainly for supernova cosmology) and the Galactic Bulge Time Domain Survey (mainly for exoplanet research with gravitational microlensing). See \citet{2025arXiv250510574O} for further details about these surveys.

From an imaging perspective, each time domain survey supplies one or more ultra deep fields. Fast {\sc Imcom} may help fully realize the potential of these ultra deep fields. By providing a unified sky atlas, it may help reduce systematic uncertainties in astrometry \citep{2019JATIS...5d4005W}. With deep, high-resolution images of nearby galaxies, it may enable new possibilities of measuring cosmic flexion \citep{2006MNRAS.365..414B} and surface brightness fluctuations \citep{1988AJ.....96..807T}. A concern is that PSFs in (Fast) {\sc Imcom} coadds are wider than native Roman PSFs; I believe this can be straightforwardly addressed in Fourier space. Similarly, deep coadds of the Galactic Bulges and Galactic Center fields may allow galactic archaeologists to detect and characterize main sequence stars (``faint objects'') in the vicinity of the galactic center.

\section{Summary} \label{sec:summ}

Image regridding and coaddition have a wide range of applications in astronomical observations. While the {\sc Imcom} algorithm \citep{2011ApJ...741...46R} has been found to meet the stringent requirements of Roman weak gravitational lensing cosmology \citep{2024MNRAS.528.2533H, 2024MNRAS.528.6680Y} and under active development and testing \citep{2025ApJS..277...55C, 2025arXiv250918286C}, its widespread usage is limited by its suboptimal efficiency. In this work, I have introduced a new algorithm, Fast {\sc Imcom}, which outperforms traditional {\sc Imcom} according to experiments in 1D; a practical implementation in 2D will be the topic of a future paper.

In Section~\ref{sec:essay}, I have laid the foundation for point spread function (PSF) manipulation. I have made the distinction between ``forward'' and ``backward'' PSFs, clarified what functions are being undersampled, and discussed the preservation of information during linear image regridding and coaddition. Then in Section~\ref{sec:meth}, I have introduced the mathematical formalisms of two specific algorithms to determine the weights, {\sc Imcom} and Fast {\sc Imcom}.

In Sections~\ref{sec:regrid} and \ref{sec:coadd}, I have systematically investigated these two algorithms in 1D. In the context of image regridding, I have demonstrated that both PSF leakage and noise amplification monotonically decrease with a wider target output PSF, and PSF residuals in Fast {\sc Imcom} results have a simple pattern. As for coaddition, I have found that Fast {\sc Imcom} is more robust than {\sc Imcom}, and that the $U$-first (prioritizing minimization of PSF leakage) and $\Sigma$-first (prioritizing minimization of noise amplification) strategies of Fast {\sc Imcom} have different advantages and disadvantages. I have also demonstrated that similar patterns are expected to apply to 2D, described the design and performance of a 2D implementation, and discussed beneficial dithering patterns.

In Section~\ref{sec:disc}, I have discussed technical challenges and potential scientific contributions of the new Fast {\sc Imcom} algorithm. I expect Fast {\sc Imcom} to facilitate investigations of error propagation, noise properties, and chromatic effects in the context of Roman weak lensing cosmology. I also believe that Fast {\sc Imcom} has great potential for other weak lensing programs, Roman time domain surveys, and beyond. I look forward to working with colleagues to realize such potential.

\section*{Acknowledgments}

I thank Christopher M. Hirata for introducing me to {\sc Imcom} and Kristen B. W. McQuinn for motivating me to pursue higher efficiency. I appreciate discussions with many people in the Roman community, especially Edward Schlafly, Henry C. Ferguson, and Arun Kannawadi.

This paper has undergone internal review in the Roman High Latitude Imaging Survey (HLIS) Cosmology Project Infrastructure Team (PIT). I would like to thank Sidney Mau and Christopher M. Hirata for helpful comments and feedback during the review process. I thank the anonymous reviewer for insightful comments and Nihar Dalal and Mike Jarvis for useful discussions during the revision.

KC received support from the ``Maximizing Cosmological Science with the Roman High Latitude Imaging Survey'' Roman Project Infrastructure Team (NASA grant 22-ROMAN11-0011).

\software{{\sc NumPy} \citep{2020Natur.585..357H}, {\sc Matplotlib} \citep{2007CSE.....9...90H}}

\section*{Data Availability}

The Fast {\sc Imcom} software is under development in the following GitHub repository:

\url{https://github.com/Roman-HLIS-Cosmology-PIT/fast_imcom.git}

The Python code and Jupyter notebooks for this paper are contained in pre-release v0.1-beta, which is available on Zenodo under an open-source Creative Commons Attribution license: \dataset[doi: 10.5281/zenodo.17938980]{https://doi.org/10.5281/zenodo.17938980}.

\appendix

\section{Impact of Asymmetric Windows} \label{app:asym}

In Section~\ref{ss:tech}, I have discussed the handling of missing pixels. In this appendix, I study a specific scenario of missing pixels: asymmetric windows for input pixels. With the Cholesky kernel of {\sc Imcom}, the selection of input pixels is unified for ${\cal O} (10^3)$ output pixels in a postage stamp (see Figure~1 of \papthree\ for an illustration), and the resulting input pixel windows can be very asymmetric for output pixels, especially those near the edge of the postage stamps.

\begin{figure*}
    \centering
    \includegraphics[width=\textwidth]{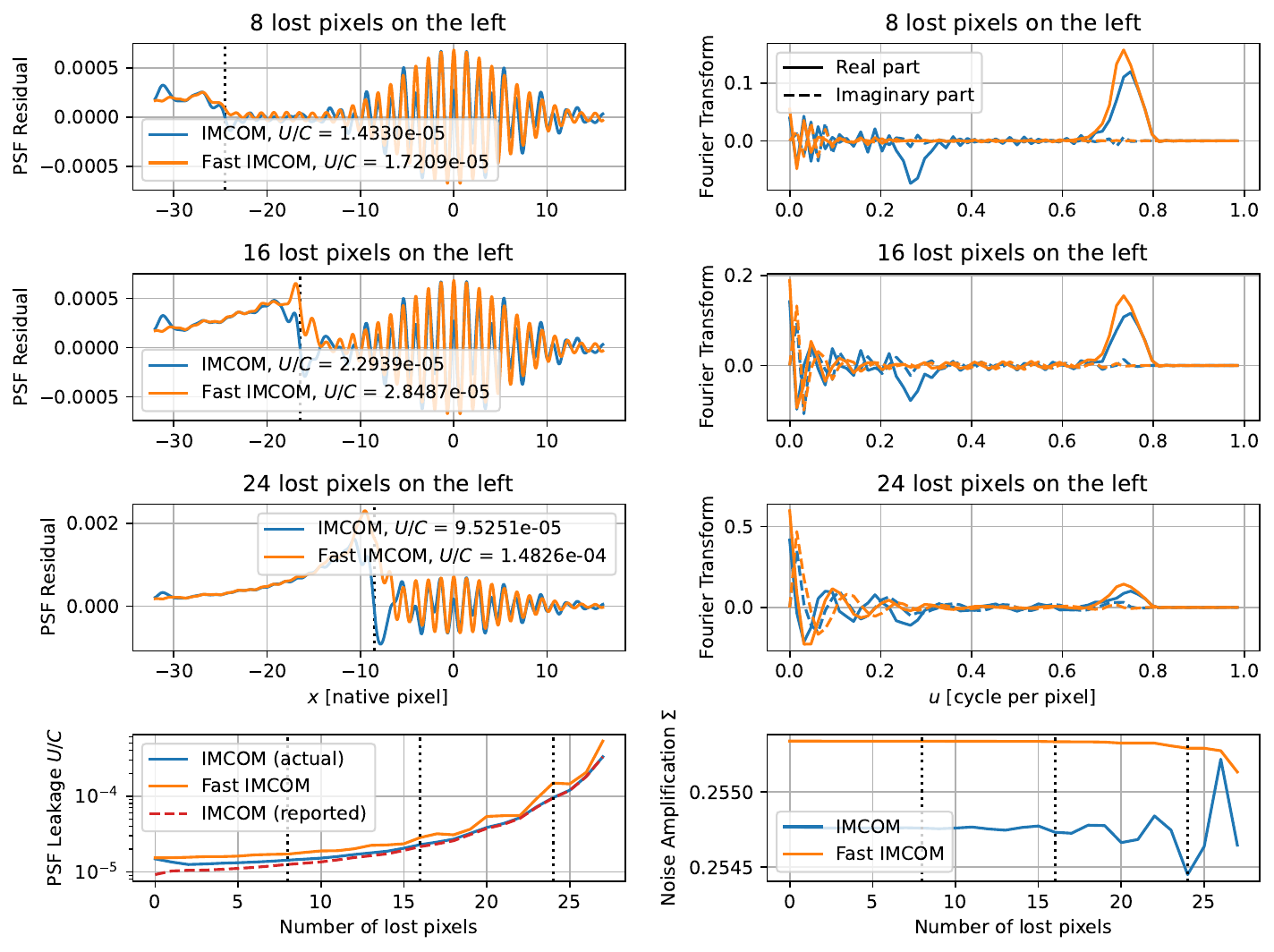}
    \caption{\label{fig:lost_pixels1}Impact of asymmetric windows on image regridding. The layout is similar to that of Figure~\ref{fig:sub_shifts1}; the difference is that the first three rows and the $x$-axes of the last row study the number of lost pixels on the left. Note that in the first three panels of the left column, the $x$-axis range is extended to show the introduced features. The edges of the asymmetric windows and the numbers of lost pixels are shown as dotted black vertical lines.}
\end{figure*}

In the 1D setup of this work (see Section~\ref{ss:setup}), I gradually remove input pixels on the left and study the impact of missing pixels on the resulting PSF residuals. Figure~\ref{fig:lost_pixels1} shows the impact on image regridding. The central wave packets of PSF residuals are basically unaffected by the asymmetric windows; meanwhile, the missing pixels introduce a smooth feature to PSF leakages on the left. With $8$ lost pixels, the peak of such feature is smaller than the amplitude of the central wave package; with $16$ lost pixels, they are comparable; with $24$ lost pixels, the new feature is more significant. The PSF fidelity deteriorates faster than exponentially as a function of the number of missing pixels.

\begin{figure*}
    \centering
    \includegraphics[width=\textwidth]{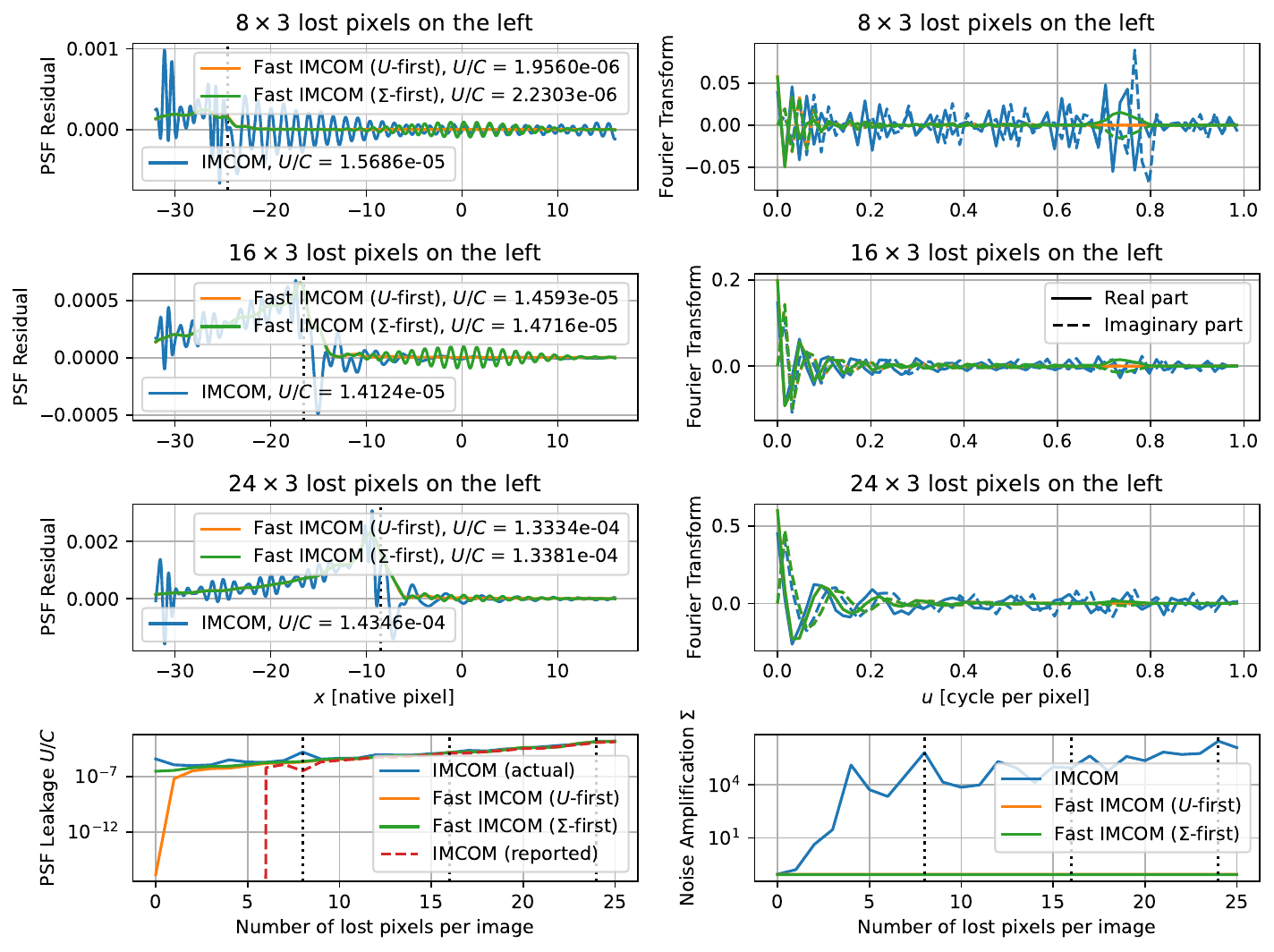}
    \caption{\label{fig:lost_pixels3}Impact of asymmetric windows on the coaddition of three images. The configuration of input images is $\Delta x_0 = 0$, $\Delta x_1 = 0.2500$ native pixels, and $\Delta x_2 = 0.6250$ native pixels. The layout is similar to that of Figure~\ref{fig:sub_shifts3}; again, the difference is that the first three rows and the $x$-axes of the last row study the number of lost pixels on the left. Like in Figure~\ref{fig:lost_pixels1}, in the first three panels of the left column, the $x$-axis range is extended to show the introduced features. The edges of the asymmetric windows and the numbers of lost pixels per image are shown as dotted black vertical lines.}
\end{figure*}

Figure~\ref{fig:lost_pixels3} shows results for the coaddition of three images. For {\sc Imcom} and both strategies of Fast {\sc Imcom}, the central wave packets are either reduced or eliminated, hence the new feature on the left seems more significant. Slightly different from the case of regridding, the PSF leakage increases exponentially with the number of lost pixels. As for noise amplification, {\sc Imcom} results are sometimes catastrophic due to numerical instabilities, while Fast {\sc Imcom} results are stable. In conclusion, wide, symmetric windows for input pixels are beneficial for PSF reconstruction. To fully leverage pixels near detector edges, one possibility is to pad the edges of input images using previous coaddition results. The exploration of such possibility is left for future work.

\bibliography{main}{}
\bibliographystyle{aasjournalv7}

\end{document}